\documentclass[amsmath,reprint,amssymb,aps,prd]{revtex4-1}

\usepackage{graphicx}% Include figure files
\usepackage{dcolumn}% Align table columns on decimal point
\usepackage{bm}% bold math

\newcommand{\avg}[1]{\langle #1\rangle}

\usepackage{hyperref} 
\begin{document}

%\preprint{APS/123-QED}

\title{Post-Newtonian Cosmological Modelling}

\author{Viraj A. A. Sanghai}
 \email{v.a.a.sanghai@qmul.ac.uk}
\author{Timothy Clifton}
 \email{t.clifton@qmul.ac.uk}
\affiliation{School of Physics and Astronomy, Queen Mary University of London, Mile End Road, London, E1 4NS, UK.\\}

%\date{\today}

\begin{abstract}
We develop a new approach to building cosmological models, in which small pieces of perturbed Minkowski space are joined together at reflection-symmetric boundaries in order to form a global, dynamical space-time. Each piece of this patchwork universe is described using post-Newtonian gravitational physics, with the large-scale expansion of the Universe being an emergent phenomenon. This approach to cosmology does not require any assumptions about non-local averaging processes. Our framework clarifies the relation between the weak-field limit of general relativity, and the cosmological solutions that result from solving Einstein's equations with a set of symmetry assumptions. It also allows the effects of structure formation on the large-scale expansion of the Universe to be investigated without averaging anything. As an explicit example, we use this formalism to investigate the cosmological behaviour of a large number of regularly arranged point-like masses. In this case we find that the large-scale expansion is well modelled by a Friedmann-like equation that contains terms that take the form of dust, radiation, and spatial curvature. The radiation term, while small compared to the dust term, is purely a result of the non-linearity of Einstein's equations.
\end{abstract}

%\pacs{98.80.Jk, 04.25.Nx}

\maketitle

\section{Introduction}

The standard approach to cosmological modelling is a top-down one in which the first step is to solve for the homogeneous and isotropic large-scale expansion. Small fluctuations on large scales are then included using first-order perturbation theory \cite{mw}, and  large fluctuations on small scales are included by appealing to Newtonian theory \cite{nbody}. This approach has many features that commend it as a good way to build cosmological models. Among the foremost of these is the mathematical simplicity involved at every step, as well as the fact that the resulting model has been found to be consistent with a wide array of cosmological observations (as long as dark matter and dark energy are allowed to be included).

Nevertheless, while the top-down approach is simple and functional, it is not necessarily self-consistent or well-defined. This is because the standard approach assumes, from the outset, that the large-scale expansion of a statistically homogeneous and isotropic universe can be accurately modelled using a single homogeneous and isotropic solution of Einstein's equations. It is far from obvious that such an assumption should be valid, as Einstein's equations are non-linear, and because it has not yet been possible to find a unique and mathematically well-defined way of averaging tensors. This makes it extremely difficult to assess the effect that the formation of structure has on the large-scale expansion of the Universe, without assuming that it is small from the outset. This is known as the â``back-reaction'' problem, which, despite much study, has uncertain consequences for the actual Universe \cite{br}.

Part of the difficulty with the investigation of the back-reaction problem is that it is hard to do cosmology without first assuming that the geometry of the Universe can be treated (to at least a first approximation) as being a homogeneous and isotropic solution of Einstein's equations. This is, unfortunately, assuming the thing that one wants to question in the first place, which is obviously not ideal. The problems with studying back-reaction in the standard top-down approach to cosmological modelling become increasingly apparent if one allows small-scale perturbations to the homogeneous and isotropic background. On small scales density contrasts must become highly non-linear, and extrapolation from the linear regime (which is assumed to be valid on large scales) can result in divergences \cite{clarkson}. On the other hand, appealing to the Newtonian theory results in a situation where the perturbations to the metric contribute terms to the field equations that are at least as large as the terms that come from the dynamical background, making the perturbative expansion itself poorly defined \cite{rasanen}.

In this paper we report on an approach that side-steps all of these difficulties. We construct cosmological models from the bottom up, by taking regions of perturbed Minkowski space, and patching them together using the appropriate junction conditions. We take the boundaries between these regions to be reflection symmetric, in order to make the problem tractable. The model that results is a space-time that is periodic, and statistically homogeneous and isotropic on large scales, while being highly inhomogeneous and anisotropic on small scales. The equations that govern the large-scale expansion of the space are determined up to post-Newtonian accuracy. The Newtonian-order equations reproduce the expected behaviour of a Friedmann-Lema\^{i}tre-Robertson-Walker (FLRW) model filled with pressureless dust, while to post-Newtonian order we find new terms that appear in the effective Friedmann equations. For the case of a single mass at the centre of each region, these terms all take the form of either dust or radiation.

With a new generation of galaxy surveys on the horizon, that have the promise to map the structure in the Universe to unprecedented levels \cite{euclid,ska}, it becomes increasingly important to understand the relativistic corrections that could be required in order to accurately interpret the data that results. To this end, higher-order corrections in perturbation theory are already being calculated (see, e.g., \cite{bonvin}). Our framework provides a way of consistently and simultaneously tracking the effects of relativistic gravity in both the regime of non-linear density contrasts, and in the large-scale cosmological expansion. A proper understanding of these effects is required to ensure we understand all possible sources of error that could arise in the interpretation of the data, and also if we are to use the data to test and understand Einstein's theory, and the dark components of the Universe.

Our models do not rely on any assumptions about the large-scale expansion being well-modelled by any homogeneous and isotropic solutions of Einstein's equations. They are also well defined on small scales, as they are explicitly constructed from the post-Newtonian expansions that are routinely used to study the weak-field and slow-motion limit of general relativity. We are therefore able to model non-linear structure within the context of a cosmological model without falling foul of any of the problems outlined above (and that appear to be inherent in any top-down approach to cosmological modelling). The large-scale expansion of our model simply emerges, as a consequence of the junction conditions. We do not have to make any assumptions about the averaging of tensors, and do not have to assume anything about the existence of any background cosmology.

Other recent approaches to bottom-up cosmological modelling include the application of geometrostatics to cosmology \cite{crt}, as well as numerical relativity techniques \cite{num, num2}, perturbative approximation schemes \cite{pert,Tim1}, and the re-discovery of the Lindquist-Wheeler models \cite{LW}. These studies have allowed the evolution of subspaces to be calculated \cite{cgrt,cgr}, numerical approximations to both the space-time and the optical properties of the space-time to be determined \cite{op1,op2,op3}, and the proof of interesting results such as the limit of many particles approaching a fluid \cite{kor1}, and the non-perturbative nature of some structures \cite{kor2,cli1}. Our study extends these previous ones, we believe, by allowing increased flexibility for the distribution of matter, while maintaining a high degree of mathematical simplicity.

In Section \ref{sec2} we introduce the post-Newtonian formalism that we will use throughout the rest of the paper. In Section \ref{sec3} we then explain how we will apply the junction conditions, in order to build a cosmological model from regions described using the post-Newtonian approximation to gravity. Section \ref{sec4} contains a detailed presentation of the field equations and the junction conditions, both expanded to post-Newtonian levels of accuracy. In Section \ref{sec5} we then manipulate these equations into a form that can be used to determine the motion of the boundaries of each of our regions, and hence the expansion of the global space-time. We then proceed, in Section \ref{sec6}, to explain how the general solution to the global expansion can be calculated, for arbitrary distributions of matter, as well as for the special case of a single mass at the centre of each region. In both cases the lowest-order Newtonian-level solution to the equations of motion give a large-scale expansion that is similar to the dust dominated homogeneous and isotropic solutions of Einstein's equations. To post-Newtonian order we find that, for the case of isolated masses, the only new terms that appear in the effective Friedmann equation look like dust and radiation. After that, in Section \ref{sec7}, we transform our solutions so that they are written as the evolution of proper distances in proper time, and on time-dependent backgrounds. We also consider the calculation of observables in these models. Finally, in Section \ref{sec8}, we conclude.

We will use Greek letters ($\mu$, $\nu$, $\rho$, ...) to denote spatial indices, and Latin letters ($a$, $b$, $c$, ...) to denote space-time indices.  We reserve the latter part of the Latin alphabet ($i$, $j$, $k$, ...) for the indices on the $2+1$-dimensional hypersurfaces that will form the boundaries of each of our regions of space. Capital Latin letters ($A$, $B$, $C$, ...) denote the spatial components of tensors on these boundaries. As usual, a comma will be used to denote a partial derivative, such that
\begin{equation}
\label{1}
\varphi_{,t} = \frac{\partial}{\partial x^{0}} \varphi  \qquad \text{and} \qquad  \varphi_{,\gamma} = \frac{\partial}{\partial x^{\gamma}} \varphi \, ,
\end{equation} 
where $x^0 = t$ here is a time coordinate, and $\varphi$ denotes any arbitrary function on space-time. Covariant derivatives will be represented by semi-colons.

\section{Post-Newtonian Formalism}  \label{sec2}

The equations of General Relativity are known to reduce to those of Newtonian gravity in the limit of slow motions ($v \ll c$) and weak gravitational fields ($\Phi \ll 1$).  In the solar system, for example, gravity is weak enough for Newton's theory to adequately explain almost all phenomena. However, there are certain effects that can only be explained using relativistic gravity. These include, for example, the shift in the perihelion of Mercury, which requires the use of relativistic gravity. To describe such situations it is useful to consider post-Newtonian gravitational physics.

The post-Newtonian formalism is essentially based on small fluctuations around Minkowski space. Both the geometry of space-time, and the components of the energy-momentum tensor, are then treated perturbatively, with an expansion parameter
\begin{equation}  \label{2}
\epsilon \equiv \frac{|\bm{v}|}{c} \ll 1, 
\end{equation} 
where $\bm{v}=v^{\alpha}$ is the  3-velocity associated with the matter fields, and $c$ is the speed of light. The first step in the post-Newtonian formalism is to associate all quantities with an ``order of smallness" in $\epsilon$. This is done for Newtonian and post-Newtonian gravitational potentials, as well as for every component of the energy-momentum tensor. 

In the remainder of this section we will outline the post-Newtonian expansion, and the quantities that are useful for solving the equations that result. Much of our discussion closely follows that of Will \cite{Will}. From this point on we will work in units where $c=1$, unless explicitly stated otherwise. 
  
\subsection{Post-Newtonian Book-Keeping}  \label{sec2a}
  
General relativity tells us that the curvature of space-time is directly related to its energy-momentum content by Einstein's equations:
 \begin{align}  \label{3}
R_{ab} = 8\pi G \left( T_{ab} - \frac{1}{2} T g_{ab} \right) \, ,
\end{align}
where $R_{ab}$ is the Ricci tensor, $g_{ab}$ is the metric of space-time, $G$ is Newton's constant, $T_{ab}$ is the energy-momentum tensor (dependent on the matter content of the Universe), and $T= g^{ab} T_{ab}$ is the trace of the energy-momentum tensor. These are a set of ten non-linear partial differential equations in four variables. 

In the vicinity of weakly gravitating systems we take the metric to be given by
\begin{equation} 
g_{ab} = \eta_{ab} + h_{ab}\ , \label{4}
\end{equation}
where $\eta_{ab}= diag(-1,1,1,1)$ is the metric of Minkowski space, and $h_{ab}$ are perturbations to that metric. We also take the energy-momentum tensor to be given by
\begin{align} \label{5}
T^{ab} = \mu u^{a} u^{b} + p ( g^{ab} + u^{a} u^{b})
\end{align}
where $\mu$ is the energy density of the matter fields measured by an observer following $u^a$,  $p$ is the isotropic pressure, and $u^{a}$ is a time-like unit 4-vector, given by
%\begin{align} \label{6}
$u^{a} = \frac{d x^{a}}{d \tau}$, 
%\end{align}
where $\tau$ is the proper time along the integral curves of $u^{a}$, and where $u^{a}$ is normalized such that $u^{a}u_{a} = -1$. Anisotropic pressure could have been included in Eq. \eqref{5}, but would only appear at $O(\epsilon^4)$ in $g_{\alpha \beta}$, and $O(\epsilon^6)$ in the equation of motion of time-like particles. This is beyond the level of accuracy used in this paper, and so we do not include it here. We can now expand $h_{ab}$, $\mu$, $p$ and $u^{a}$ in orders of $\epsilon$, and relate the resultant quantities to each other via Eq. \eqref{3}.
 
To begin this we first note that, in the post-Newtonian formalism, time derivatives add an extra degree of smallness to the object they operate on, as compared to spatial derivatives. This follows because the time variations of the metric and energy-momentum tensors are taken to be a result of the motion of the matter in the space-time, such that  
\begin{equation} \label{7}
\varphi_{,t} \sim |\bm{v}| \ \varphi_{,\gamma} \ ,
\end{equation} 
where $\bm{v}$ is the 3-velocity of the matter fields, and $\varphi$ is any space and time dependent function in the system (such as one of the components of $h_{ab}$ or $T_{ab}$).  

To find the lowest-order part of $h_{tt}$ we note that the leading-order part of the equation of motion for a time-like particle takes the same form as in Newtonian theory. That is,
%\begin{equation}  \label{8}
$u^{\gamma}_{\ ,t} = \frac{1}{2}h_{tt,\gamma}  $.
%\end{equation} 
As $u^{\gamma} \sim {\epsilon}$, we therefore have that the leading-order part of $h_{tt}$ is 
\begin{equation}  \label{9}
h_{tt} \sim {\epsilon^2} \ . 
\end{equation}
Similar considerations lead to the conclusion that the leading-order part of the spatial components of the metric are given by
\begin{equation}
h_{\alpha \beta} \sim \epsilon^2 \, ,
\end{equation}
while those of the $t\alpha$-components are given by
\begin{equation}
h_{t \alpha} \sim \epsilon^3 \, .
\end{equation}
The next-to-leading-order parts of each of these components is $O(\epsilon^2)$ smaller than the leading-order part, in every case.

Similarly, the lowest-order part of $\mu$ can be determined from the leading-order part of the $tt$-component of Eq. \eqref{3}. This takes the form of the Newton-Poisson equation,
%\begin{equation}  \label{10}
$h_{tt,\gamma\gamma}  = - 8\pi G \mu  $, 
%\end{equation}
so that the lowest-order part of $\mu$ can be seen to be 
\begin{equation}  \label{11}
\mu \sim \epsilon^2 \ .
\end{equation}
Here, and throughout, we have chosen units such that spatial derivatives do not change the order-of-smallness of the object on which they operate. To find the lowest-order contribution to the pressure we can consider the conservation of energy-momentum, $T^{ab}_{\quad ; b} = 0$. The lowest-order part of the spatial component of these equations is $\mu (u^{\alpha})_{,t} + \mu u^{\beta} (u^{\alpha})_{,\beta} = \frac{1}{2} \mu h_{tt,\alpha} - p_{,\alpha} \label{16}$, 
from which it can be seen that
\begin{equation}
p \sim  \epsilon^4 \, .
\end{equation}
Again, the next-to-leading-order part of the energy density is $O(\epsilon^2)$ smaller than the leading-order part, while the higher-order corrections to the pressure will not be required for what follows.
 
\subsection{Post-Newtonian Potentials} \label{sec2b}

In order to solve the equations of both Newtonian and post-Newtonian gravitational physics it is useful to define some potentials, as well as make some identifications for the components of the energy-momentum tensor. The first of these involves the leading-order part of the energy density, which we write as
\begin{equation}
\mu^{(2)} = \rho \, ,
\end{equation}
where $\rho$ is the density of mass. Here, and throughout, a superscript in brackets denotes a quantity's order-of-smallness in $\epsilon$. The next-to-leading order part of the energy density is then written as
\begin{equation}
\mu^{(4)} = \rho \Pi \, ,
\end{equation}
where $\Pi$ is known as the the specific energy density. In what follows, we will also use $v^{\alpha}$ to denote the spatial components of $u^a$ at lowest order.

Using $\rho$ we can define the first of our potentials, which is simply the Newtonian  gravitational potential, defined implicitly as the solution to 
\begin{equation} \label{13} 
\nabla^2 \Phi \equiv -4\pi G \rho \, ,
\end{equation}
where $\nabla^2 = \partial_{\alpha} \partial_{\alpha}$ is the 3-dimensional Laplacian operator of flat space. The reason for defining the potential in this way is very simple: it allows us to write the solution to $h^{(2)}_{tt,\gamma\gamma}  = - 8\pi G \mu$ as $h^{(2)}_{tt} = 2 \Phi$. At this point the ``solution'' for $h_{tt}$ is little more than a change of notation (with obvious historical significance). When it comes to post-Newtonian potentials, however, the equations become much more complicated. This change of notation is then much more useful, especially if the potentials that we define are simply solutions to Poisson's equations.

With this in mind, it is useful to make the following implicit definitions for new gravitational potentials
\begin{align}  
\nabla^2 \chi &\equiv -2\Phi  \, , \nonumber \\
\nabla^2 V_{\mu} &\equiv -4\pi G \rho v_{\mu} \, , \nonumber \\
\nabla^2 \Phi_{1}  &\equiv - 4\pi G \rho v^2 \, , \nonumber \\
\nabla^2 \Phi_{2} &\equiv  - 4\pi G \rho \Phi \, ,\nonumber \\
\nabla^2 \Phi_{3}  &\equiv - 4\pi G \rho \Pi \, , \nonumber \\
\nabla^2 \Phi_{4}  &\equiv - 4\pi G p \, , \nonumber
\end{align} 
where $v^2 = v^{\alpha}v_{\alpha}$, $\chi \sim \epsilon^2$, $V_{\mu} \sim \epsilon^3$ and $\Phi_{1} \sim \Phi_{2} \sim \Phi_{3}\sim \Phi_{4} \sim  \epsilon^4$. In what follows, we will not require any potentials of order higher than $\epsilon^4$. 
 
\subsection{Green's Functions} \label{sec2c}

When applying the post-Newtonian formalism to isolated gravitational systems, such as the Sun, it is usual to assume that the space-time is asymptotically flat. This allows the boundary terms of Green's functions to be neglected, so that the solutions to Poisson's equation are given by simple integrals over spatial volumes. This simplicity is a substantial benefit when solving for the potentials defined in the equations above. In the case of a cosmological model, however, we cannot assume asymptotic flatness. We must, therefore, be more careful with the boundary terms.

The equations we want to solve take the form of the Poisson equation,
\begin{align} \label{96}
\nabla^2 \varphi = \mathcal{F} \,, 
\end{align}
where $\varphi$ is now being used to denote a potential, and where $\mathcal{F}$ is some function on space-time (such as the mass density). To solve this equation we consider the Green's function, $\mathcal{G}(\mathbf{x}, \mathbf{x'}, t)$, that satisfies 
\begin{align} \label{98}
\nabla^2 \mathcal{G}= -\delta(\mathbf{x-x'}) + C_{1} \, ,
\end{align}
where $\mathbf{x}$ and $\mathbf{x'}$ denote spatial positions, where $\delta(\mathbf{x-x'})$ is the Dirac delta function, and where $C_{1}$ is a constant over spatial hypersurfaces (the need for $C_{1}$ in this equation will become apparent shortly). 

We want to solve Eq. \eqref{96} over a spatial volume $\Omega$, with boundary $\partial \Omega$. The Green's function we will use for this must, of course, satisfy Gauss' theorem on this domain, such that
\begin{align} \label{100}
\int_{\Omega}  \nabla^2 \mathcal{G} \ dV = \int_{\partial \Omega}  \bm{n} \cdot \nabla \mathcal{G} \ dS \, ,
\end{align} 
where $\bm{n}$ is the unit-vector normal to the boundary. If we now choose $\bm{n} \cdot \nabla \mathcal{G} |_{\partial \Omega} = 0$ as the boundary condition for $\mathcal{G}$, then Eqs. \eqref{98} and \eqref{100} imply
\begin{align} \label{101}
C_{1} = \frac{1}{V}  \ , 
\end{align} 
where $V$ is the spatial volume of the cell. The solution to Eq. \eqref{96} is then given, in terms of $\mathcal{G}$, by considering
\begin{align} 
\int_{\Omega} \mathcal{G} \mathcal{F} \ dV &= \int_{\Omega} \mathcal{G}\nabla^2 \varphi \ dV \nonumber \\
&= \int_{\Omega} [\nabla\cdot (\mathcal{G} \nabla \varphi) - \nabla \mathcal{G} \cdot \nabla \varphi] \ dV\nonumber \\
&= \int_{\Omega} [\nabla \cdot (\mathcal{G} \nabla \varphi) -\nabla \cdot (\varphi \nabla \mathcal{G}) + \varphi \nabla^2 \mathcal{G}] \ dV\nonumber  \\
&= \int_{\Omega} \nabla \cdot (\mathcal{G} \nabla \varphi-  \varphi \nabla \mathcal{G}) \ dV - \varphi + \bar{\varphi} \, , \nonumber 
\end{align}
where $\bar{\varphi} = C_{1}\int_{\Omega} \varphi \ dV = \frac{1}{V} \int_{\Omega} \varphi \ dV$ is a constant over $\Omega$. Rearranging, the potential $\varphi$ can be seen to be given by
\begin{align} \label{102}
\varphi &= \bar{\varphi} - \int_{\Omega} \mathcal{G}  \mathcal{F} \ dV + \int_{\partial \Omega} \mathcal{G} \bm{n} \cdot  \nabla \varphi \ dA \, , 
\end{align} 
where we have again made use of the boundary condition $\bm{n} \cdot \nabla \mathcal{G} |_{\partial \Omega} = 0$.

If one were now to assume that $\varphi$ was asymptotically flat, then the first and last terms on the right-hand side of Eq. \eqref{102} would vanish. The Green's function would then take the form of the Newton kernel, such that
\begin{equation} \label{18}
\varphi(\mathbf{x},t) =  -\frac{1}{4 \pi} \int_{\Omega}{\frac{ \mathcal{F}}{ |\mathbf{x} - \mathbf{x'}| } d^3 x'} \, .
\end{equation}
However, these assumptions are not true in general, and especially not in the case of cosmological modelling. This is because cosmological models do not have asymptotically flat regions, by their very definition. In what follows, we must therefore be more careful. We have to specify appropriate boundary conditions for our potentials, and include the boundary terms in Eq. \eqref{102}, if we are to use the Green's function formalism to determine the form of our gravitational potentials.
 
\section{BUILDING A COSMOLOGY USING JUNCTION CONDITIONS} \label{sec3}

In this paper we will take a bottom-up approach to cosmological modelling.  This will involve considering cosmological models that are constructed from large numbers of cells, that can be put next to each other to form a periodic lattice structure. The shape of each cell will be taken to be a regular polyhedron, and will be assumed to be identical to every other cell, up to rotations, reflections and translations. 

\begin{figure}
    \includegraphics[width=85mm]{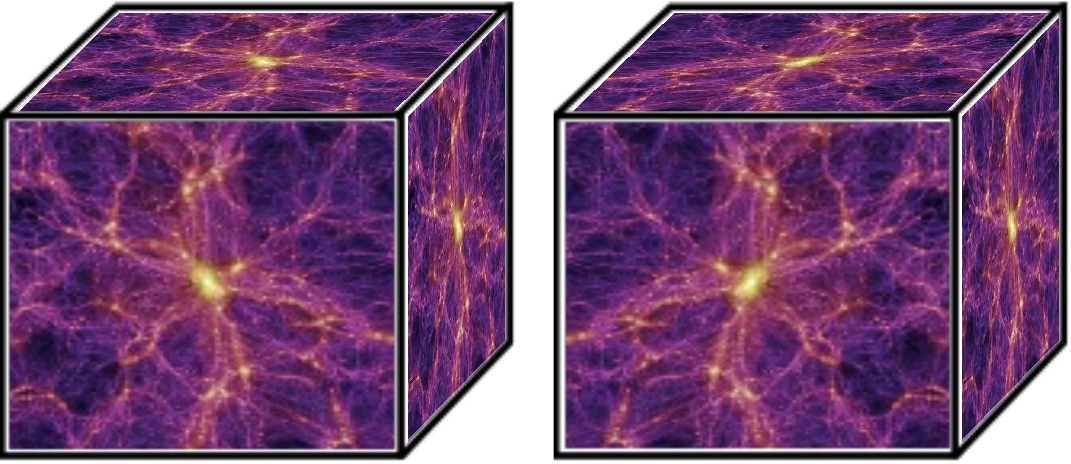}
      \caption{\label{fig1} Two adjacent cubic cells, with example matter content, consisting of filaments and voids. The second cell is the mirror image of the first. This figure was produced using an image from \cite{volk}.} 
   \end{figure}

The physical systems that we intend to model with these cells will depend on the size of cell that we are considering. For example, for cells that are approximately the size of the homogeneity scale (about 100 Mpc), we could consider modelling clusters of galaxies, as illustrated in Fig. \ref{fig1}. Other systems, such as individual galaxies, could equally well be modelled with cell sizes of the order of about 1 Mpc. The only requirement we have is that the system must satisfy the requirements of the post-Newtonian formalism. Specifically, this means that $v\ll c$ and $p \ll \rho$, so that the bulk of the interior of each cell is described by Newtonian and post-Newtonian gravitational physics. 

The post-Newtonian formalism is expected to work well in the regime of non-linear density contrasts, and so should be expected to be adequate for modelling most aspects of the gravitational fields of galaxies and clusters of galaxies. This formalism is, however, limited to scales much smaller than the cosmological horizon. We therefore require each of our cells to be much smaller than the Hubble radius, $H_{0}^{-1}$. A violation of this requirement would result in matter at the boundary of a cell moving at close to the speed of light.  We also assume each cell is filled with normal matter, so that pressures are small with respect to energy densities.

We note that the post-Newtonian framework cannot, and should not, be used to describe multiple cells simultaneously. However, due to the periodicity of our lattice structure, we only need to know the space-time geometry of any one cell, and its boundary conditions with neighbouring cells. As we will show below, this information is sufficient to tell us how we should expect the entire Universe to evolve. 

Let us now turn to a more detailed consideration of the conditions required in order to join two cells together at a boundary. First and foremost, the cells must satisfy certain smoothness requirements across their respective boundaries, known as the Israel junction conditions, if their union is to be a solution to Einstein's equations. These conditions, in the absence of surface layers, are given by \cite{Is1}
\begin{align} 
[\gamma_{ij}] &= 0 \,  \label{gamma} \\  
[{K}_{ij}] &=0 \, , \label{19}
\end{align}
where $[\varphi] = \varphi^{(+)} - \varphi^{(-)}$ denotes the jump across the boundary for any quantity $\varphi$, and the $i$ and $j$ indices denote tensor components on the boundary. The $^{(+)}$ and $^{(-)}$ superscripts here show that a quantity is to be evaluated on either side of the boundary (i.e. on the sides labelled by $+$ or $-$, respectively). 

In these equations, $\gamma_{ij}$ is the induced metric on the boundary, and ${K}_{ij}$ is the extrinsic curvature of the boundary, defined by
\begin{align} \label{20a}
\gamma_{ij} &\equiv \frac{\partial{x^{a}}}{\partial{\xi^{i}}}\frac{\partial{x^{b}}}{\partial{\xi^{j}}} g_{ab} 
\end{align}
and
\begin{align}
{K}_{ij}  &\equiv  \frac{\partial{x^{a}}}{\partial{\xi^{i}}}\frac{\partial{x^{b}}}{\partial{\xi^{j}}} n_{a;b} \ , \label{20}
\end{align}
where $\xi^{i}$ denotes the coordinates on the boundary, and $n^{a}$ is the space-like unit vector normal to the boundary.

\begin{figure}
\includegraphics[width=85mm]{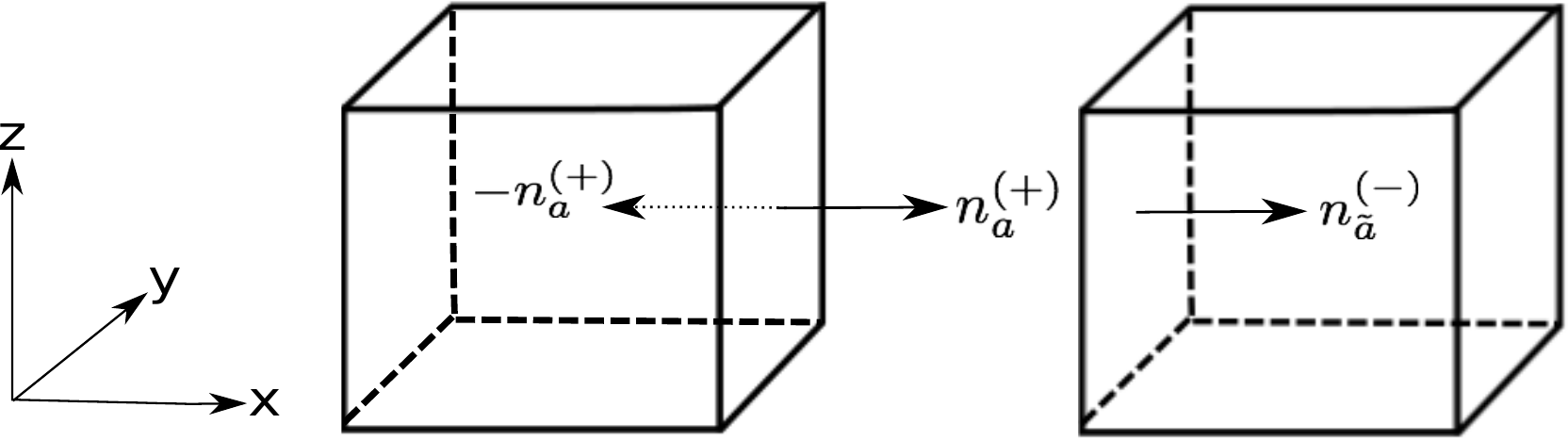}
\caption{\label{fig2} A schematic diagram showing the normal vectors involved in the junction conditions, $n^{(+)}_{a}$ and $n^{(-)}_{\tilde{a}}$. The vector $-n^{(+)}_{a}$ is shown as a dashed arrow, and is the mirror image of $n^{(-)}_{\tilde{a}}$.} 
\end{figure}

In our construction we choose to consider reflection symmetric boundaries. The Israel junction conditions can then be simplified. The situation we wish to consider is illustrated in Fig. \ref{fig2}, for two cubic cells. We use $x^{a}$ and $x^{\tilde{a}}$ to denote the coordinates used within the first and second cells, respectively. Reflection symmetry means that Eq. \eqref{gamma} is automatically satisfied. The second junction condition, given by Eq. \eqref{19}, can be written as
\begin{align}  \label{21}
\frac{\partial{x^{a}}}{\partial{\xi^{i}}}\frac{\partial{x^{b}}}{\partial{\xi^{j}}} n^{(+)}_{a;b} =  \frac{\partial{x^{\tilde{a}}}}{\partial{\xi^{i}}}\frac{\partial{x^{\tilde{b}}}}{\partial{\xi^{j}}} n^{(-)}_{\tilde{a};\tilde{b}}\ ,
\end{align}
where $n^{(+)}_{a}$ and  $n^{(-)}_{\tilde{a}}$ are outward and inward pointing normals, in the first and second cells, respectively. They are shown in Fig. \ref{fig2}. Now, mirror symmetry implies that $n^{(-)}_{\tilde{a}}= -n^{(+)}_{a}$. Symmetry therefore demands that
\begin{align}  \label{22}
\frac{\partial{x^{a}}}{\partial{\xi^{i}}}\frac{\partial{x^{b}}}{\partial{\xi^{j}}} n^{(+)}_{a;b} =  -\frac{\partial{x^{\tilde{a}}}}{\partial{\xi^{i}}}\frac{\partial{x^{\tilde{b}}}}{\partial{\xi^{j}}} n^{(+)}_{\tilde{a};\tilde{b}} \, .
\end{align}
This implies that ${K}_{ij} = - {K}_{ij}$, or, in other words, that the extrinsic curvature must vanish on the boundary of every cell, i.e.
\begin{equation}  \label{23}
{K}_{ij}=0 \, .
\end{equation}
This equation must be satisfied on each and every boundary in our lattice.

\renewcommand{\arraystretch}{1.35}

\begin{table}[b]
\begin{center}
\begin{tabular}{|c|c|c|c|}
\hline
$\begin{array}{c} \bf{Lattice}\\
  \bf{Structure} \end{array}$ & $\begin{array}{c} \bf{Lattice}\\
  \bf{Curvature} \end{array}$ & $\begin{array}{c} \textbf{Cell}\\
  \textbf{Shape} \end{array}$
& $\begin{array}{c} \textbf{Cells per}\\
  \textbf{Lattice} \end{array}$ \\ 
\hline
\{333\} & + & Tetrahedron & 5 \\
\{433\} & + & Cube & 8 \\
\{334\} & + & Tetrahedron & 16 \\
\{343\} & + & Octahedron & 24 \\
\{533\} & + & Dodecahedron & 120 \\
\{335\} & + & Tetrahedron & 600 \\
\{434\} & 0 & Cube & $\infty$ \\
\{435\} & - & Cube & $\infty$ \\
\{534\} & - & Dodecahedron & $\infty$ \\
\{535\} & - & Dodecahedron & $\infty$ \\
\{353\} & - & Icosahedron & $\infty$\\
\hline
\end{tabular}
\end{center}
\caption{{\protect{\textit{A summary of all regular lattice structures that can exist on 3-surfaces of constant curvature. Hyper-spherical lattices are denoted by $+$, flat lattices by $0$, and hyperbolic lattices by $-$.  The lattice structure is given by the Schl\"{a}fli symbols, \{$pqr$\}, which are explained in the text.  The shape of the cells, and the number of cells in the lattice are also given. For further details of these structures see} \cite{poly}.}}}
\label{table1}
\end{table}

Eq. \eqref{23} is valid for any cell shape, as long as the boundaries are reflection symmetric. In general, however, there are only a finite number of ways that spaces of constant curvature can be tiled with regular convex polyhedra. These are listed in Table \ref{table1}, where we have also included the Schl\"{a}fli symbols that allow the structure of the lattice to be inferred, and the number of cells in each of the different structures. A Schl\"{a}fli symbol $\{ p q r \}$ corresponds to a lattice with $p$ edges to every cell face, $q$ cell faces meeting at every vertex of every cell, and $r$ cells meeting around every cell edge. One may note that regular lattices on hyper-spherical spaces have a maximum of $600$ cells, while those on flat and hyperbolic lattices always have an infinite number of cells.

In what follows we will often require knowledge of the total surface area of a cell, $A$. For each of the five different polyhedra in Table \ref{table1} we can write $A= \alpha_{\kappa} X^2$, where $X$ is the distance from the centre of a cell to the centre of a cell face, and where $\kappa$ denotes the number of faces per cell (e.g. a cube has $\kappa=6$ faces). For the purposes of creating the diagram in Fig. \ref{fig2}, we chose to consider cube-shaped cells. In what follows we will also often consider this particular case. One of the advantages of tessellating the Universe into cubic cells is that tilings of this type exist for open, closed and flat universes. Another advantage is that Cartesian coordinates can be used, and aligned with the symmetries of the cells.

Let us now finish this section by briefly considering the motion of a boundary at $x = X(t,y,z)$, where the $x$-direction has been chosen to be orthogonal to the centre of a cell face (as in Fig. \ref{fig2}). The $4$-velocity of this boundary is then given by the following expression
\begin{align} \label{24}
U^{a} \equiv \frac{dx^{a}}{d\tau} = \frac{dt}{d\tau} \bigg(1; \frac{dX}{dt},0,0\bigg)  \ ,
\end{align}
where $x^{a}$ are the coordinates of points on the boundary, where $\tau$ is the proper time measured along a time-like curve in the boundary, and where we have chosen the integral curves of $U^a$ to stay at fixed $y$ and $z$ coordinates. This vector is orthogonal to the space-like normal to the cell face, such that $U^{a}n_{a}=0$.

\begin{table}[t]
\begin{tabular}{ | c | c | c | }
\hline
$\begin{array}{c} \textbf{Cell}\\
  \textbf{Shape} \end{array}$ & $\begin{array}{c} \bf{Faces \, per}\\
  \bf{Cell, \,} {\bm \kappa} \end{array}$ & $\begin{array}{c} \textbf{Surface \, Area}\\
  \textbf{Coefficients, \,} {\bm \alpha_{\kappa}} \end{array}$\\ 
     \hline
    Tetrahedron & 4 & $24\sqrt{3}$   \\ %\hline %radius of insphere X - a\ \sqrt{24}
    Cube & 6 & 24 \\  %\hline
    Octahedron & 8 & $12\sqrt{3}$  \\ %\hline
    Dodecahedron & 12 & $120\frac{\sqrt{25 + 10\sqrt{5}}}{25 + 11\sqrt{5}}$ \\ %\hline
    Icosahedron & 20 & $\frac{120\sqrt{3}}{7 + 3\sqrt{5}}$ \\ \hline
\end{tabular}
\caption{\label{tab} The five possible different cell shapes, together with the number of faces per cell, and the numerical coefficients for the surface area, $\alpha_{\kappa} \equiv A/X^2$.} 
\end{table}

We can now define two types of derivatives along the boundary, in time-like and space-like directions, respectively. These are given to lowest order by
\begin{align} \label{25}
 ^{.}  &\equiv U^{a} \partial_{a} \equiv \partial_{t} + X_{,t} \partial_{x} \nonumber \\ 
 _{|A}&\equiv m^{a} \partial_{a} \equiv \partial_{t} + X_{,A} \partial_{x}\ ,  
\end{align}
where $m^{a}$ is a space-like vector in the cell face, which satisfies $m^{a}n_{a}=0$. These expressions allow us to write the lowest-order parts of $n_t$ and $n_A$ as
\begin{align}  \label{26}
n_{t} &= -n_{x} X_{,t}  \nonumber \\ 
n_{A} &= -n_{x} X_{,A}\ . 
\end{align}
We know that $X_{,t} \sim \epsilon$ and $n_{x} \sim 1$, which means that the leading-order part of $n_{t} \sim \epsilon$. This information will be used below.

\section{Governing Equations} \label{sec4}

In this section we will present the equations that govern the dynamics of each cell, and its matter content. We will first use Einstein's field equations to relate the space-time metric (given in Eq. \eqref{4}) to the energy-momentum content of the cell (given in Eq. \eqref{5}). After this, we will evaluate the extrinsic curvature of the cell boundaries, using the same geometry. The extrinsic curvature will be required to vanish, in order to satisfy the reflection symmetric boundary conditions, and will provide us with the information required to study the evolution of the boundary.

\subsection{Einstein's Field Equations} \label{sec4a}

To begin this study, we need to evaluate the Ricci tensor for the perturbed Minkowski space given in Eq. \eqref{4}. Recall that the leading-order contributions to the metric perturbations are at $h_{ t t} \sim h_{\mu\nu} \sim \epsilon^2$, and  $h_{t\mu} \sim \epsilon^3$. The leading-order components of the Ricci tensor are then given by
\begin{align} \label{27}
R^{(2)}_{ t t}  &=  - \frac{\nabla^2 h^{(2)}_{ t t}}{2} \,  ,  \\  
R^{(2)}_{\mu\nu}  &= \frac{1}{2} [h^{(2)}_{\mu\alpha,\nu\alpha} + h^{(2)}_{\nu\alpha,\mu\alpha} - h^{(2)}_{\mu\nu,\alpha\alpha} %\nonumber \\ &\, \hspace{2.5cm}
- h^{(2)}_{\alpha\alpha,\mu\nu} +h^{(2)}_{ t t,\mu\nu}] \, ,  \label{30} 
\end{align}
and
\begin{align}
\label{29}
R^{(3)}_{ t\mu}  &=  - \frac{1}{2}[-h^{(2)}_{\mu\nu, t\nu} + h^{(2)}_{\nu\nu, t\mu} + h^{(3)}_{ t\mu,\nu\nu} - h^{(3)}_{ t\nu,\nu\mu}]  \, . 
\end{align}
Here we have used the short-hand notation $\nabla^2 = \partial_{\mu}\partial_{\mu}$, and $|\nabla h^{(2)}_{ t t}|^2 = h^{(2)}_{ t t,\alpha}h^{(2)}_{ t t,\alpha}$. As before, we have chosen units such that spatial derivatives do not add an order of smallness, and have used super-scripts in brackets to denote the order of a quantity. 

The only higher-order part of the Ricci tensor that we require will be the $O(\epsilon^4)$ part of the $tt$-component, which is given by
\begin{align}  
R^{(4)}_{ t t}  &=   \frac{1}{2} \bigg[2h^{(3)}_{ t\nu,\nu t} -\frac{1}{2}|\nabla h^{(2)}_{ t t}|^2 - \nabla^2 h_{ t t}^{(4)} - h^{(2)}_{\mu\mu, t t}  \label{28} \\
& \hspace{2cm} + h^{(2)}_{ t t,\nu} ( h^{(2)}_{\nu\alpha,\alpha} - %factor of 1/2 missing
\frac{1}{2} h^{(2)}_{\alpha\alpha,\nu}) + \frac{1}{2} h^{(2)}_{\nu\alpha} h^{(2)}_{ t t,\nu\alpha}\bigg] \, . \nonumber
\end{align}
No other components of the Ricci tensor will be required at this order.

To proceed further we now need to make a gauge choice, in order to eliminate superfluous degrees of freedom. For this we use the standard post-Newtonian gauge, which is given by \cite{Will}
\begin{align} \label{31} 
& \frac{1}{2} h^{(2)}_{ t t,\mu} + h^{(2)}_{\mu\nu, \nu} - \frac{1}{2} h^{(2)}_{\nu \nu,\mu} =  0 \, , 
%\quad (\text{at} \  O(\epsilon^2)) 
\end{align}
and
\begin{align} \label{32}
& h^{(3)}_{\nu t, \nu} - \frac{1}{2} h^{(2)}_{\nu\nu, t} =  0 \,   .
%\quad (\text{at} \  O(\epsilon^3))  
\end{align}
Note that this is not the same as the harmonic gauge.

The perturbed metric, Eq. \eqref{4}, and the definition of the 4-velocity can now be used to write the 4-velocity as
 \begin{align} \label{33}
u^{a} =  \bigg(1 +\frac{h^{(2)}_{tt}}{2}+ \frac{v^2}{2}\bigg)(1;v^{\mu}) +  O(\epsilon^4) \ ,
 \end{align} 
where $v^2 = v^{\mu}v_{\mu}$. We can use this equation to give us the components of $T_{ab}$, up to post-Newtonian levels of accuracy. We note that in order to evaluate the field equations at $O(\epsilon^2)$ we only need to know $T_{ t t} = -T = \rho$.
Using Eqs. \eqref{27} and \eqref{30}, and the gauge conditions \eqref{31} and \eqref{32}, the field equations \eqref{3} then give us
 \begin{align} \label{34}
\nabla^2 h^{(2)}_{ t t} &= -8 \pi G \rho + O(\epsilon^4)\ ,  \\
\nabla^2 h^{(2)}_{\mu\nu} &=  -8 \pi G \rho \delta_{\mu\nu}+ O(\epsilon^4) \ . \label{35}
\end{align}
Using the potentials defined in Section \ref{sec2}, we then find
\begin{align} \label{36} 
& h^{(2)}_{tt} = 2\Phi \, , 
\end{align}
and
\begin{align}
& h^{(2)}_{\mu\nu} = 2\Phi \delta_{\mu\nu} \, .   \label{37}
\end{align}

These solutions, together with our gauge conditions \eqref{31} and \eqref{32}, allow us to simplify Eq. \eqref{29}, to get
\begin{align} \label{38}
R^{(3)}_{ t\mu}  &=  -\frac{1}{2} \bigg[  h^{(3)}_{ t\mu,\nu\nu} + \Phi_{, t\mu} \bigg] \, . 
\end{align}
The field equations \eqref{3} then give
\begin{align} \label{39}
& h^{(3)}_{ t\mu,\nu\nu} + \Phi_{, t\mu}   =  16 \pi G \rho v_{\mu} \, ,
\end{align}
which has the solution
\begin{align}  \label{40}
 h^{(3)}_{ t\mu} = -4V_{\mu} + \frac{1}{2} \chi_{, t\mu} \, ,  
\end{align}
where we have again made use of the potentials defined in Section \ref{sec2}. Eqs. \eqref{36}, \eqref{37} and \eqref{40} give the leading-order contributions to all of the components of the perturbed metric.

To go further, we now need to evaluate $h_{ t t}^{(4)}$. This will be done using Eq. \eqref{28}, our gauge conditions \eqref{31} and \eqref{32}, and the lowest-order solutions found above. The relevant part of the Ricci tensor then simplifies to
\begin{align}  \label{41}
R^{(4)}_{ t t} &=   \frac{1}{2} \bigg[ -4|\nabla \Phi|^2 - \nabla^2 h_{ t t}^{(4)} + 4 \Phi \nabla^2 \Phi \bigg] \, ,
\end{align}
which, using the identity $|\nabla \Phi|^2 = \frac{1}{2} \nabla^2 \Phi^2 - \Phi \nabla^2 \Phi$, can be written as
\begin{align}  \label{42}
R^{(4)}_{ t t} &=   -\frac{1}{2} \bigg[ \nabla^2 (2\Phi^2) - 8\Phi \nabla^2 \Phi  + \nabla^2 h_{ t t}^{(4)} \bigg] \, .
\end{align} 
Similarly, we can write the $tt$-component of the right-hand side of Eq. \eqref{3} as
\begin{align}  \label{43}
 T_{ t t} - \frac{1}{2} Tg_{tt} = \rho\bigg(v^2 - \Phi + \frac{1}{2} \Pi + \frac{3}{2} \frac{p}{\rho}\bigg) + O(\epsilon^6) \ .
\end{align} 
Equating Eqs. \eqref{42} and \eqref{43}, and using the field equations \eqref{3}, we then find that
\begin{align}   \label{44}
 h_{ t t}^{(4)} = -2\Phi^2 + 4 \Phi_{1} + 4 \Phi_{2} + 2 \Phi_{3} + 6 \Phi_{4} \, . 
\end{align}
Once more, this solution has been written in terms of the potentials defined in Section \ref{sec2}. Eqs. \eqref{36}, \eqref{37}, \eqref{40} and \eqref{44} give all of the components of the metric that we will require.

\subsection{Extrinsic Curvature Equations} \label{sec4b}

Let us now calculate the extrinsic curvature of the boundary. To do this we require the covariant derivative of the normal, $n_{a;b}$. As stated earlier, the leading order parts of $n_{t}$ and $n_{\mu}$ are $O(\epsilon)$ and $O(1)$, respectively. The components of $n_{a;b}$ can then be seen to be given, up to the required order, by
\begin{align}  \label{45}
n_{ t ; t}  &=  n_{ t, t} + \frac{h^{(2)}_{ t t,\mu}n_{\mu}}{2} \\&\quad- n_{\mu} \bigg[ \frac{h^{(2)}_{\mu\nu} h^{(2)}_{ t t,\nu}}{2}  -\frac{h^{(4)}_{ t t,\mu}}{2} + h^{(3)}_{ t\mu, t}  \bigg]  \nonumber 
 +\frac{h^{(2)}_{ t t, t} n_{ t}}{2} + O(\epsilon^6) \, ,  
\end{align}
and
\begin{align} \label{46}
n_{ t ;\mu}  &=  n_{ t,\mu}  \\
& \quad+ \frac{h^{(2)}_{ t t,\mu} n_{ t}}{2} - \frac{n_{\nu}}{2} \bigg[ h^{(2)}_{\mu\nu, t} - h^{(3)}_{ t\mu,\nu} +  h^{(3)}_{ t\nu,\mu}  \bigg] \nonumber  + O(\epsilon^5) \, , 
\end{align}
and
\begin{align} \label{47}
n_{\mu ; \nu}  &=   n_{\mu,\nu} \\&\quad- \frac{1}{2} n_{\alpha}  (-h^{(2)}_{\mu\nu,\alpha} + h^{(2)}_{\alpha\mu,\nu} +  h^{(2)}_{\alpha\nu,\mu} )   + O(\epsilon^4) \, . \nonumber
\end{align}
New lines have been used, in each of these equations, to separate terms of different orders.

The extrinsic curvature of the cell boundaries can now be calculated using Eq. \eqref{20}. The $tt$-component of this equation is given, to lowest order, by
\begin{align}  \label{49}
{K}^{(2)}_{tt}&=  -n_{x}X_{,tt}+\frac{h^{(2)}_{ t t,\mu} n_{\mu}}{2} \, , 
\end{align}
where we have used $n_{ t} = -n_{x} X_{,t}$. At next-to-leading-order have
\begin{align} \label{50}
{K}^{(4)}_{tt} &= \frac{X_{,t}^{2}}{2}  n_{\alpha}  (h^{(2)}_{xx,\alpha} - 2h^{(2)}_{\alpha x,x} ) + \frac{h^{(2)}_{ t t,\mu} n^{(2)}_{\mu}}{2}-n^{(2)}_{x}X_{,tt}  \nonumber \\  
&\quad - 2X_{,t}\bigg[ \frac{h^{(2)}_{ t t,x} n_{x}X_{,t}}{2} + \frac{n_{\nu}}{2} ( h^{(2)}_{x\nu, t} - h^{(3)}_{ tx,\nu} +  h^{(3)}_{ t\nu,x}  ) \bigg]  \nonumber \\
&\quad -\frac{h^{(2)}_{ t t, t} X_{,t}}{2} - n_{\mu} \bigg[ \frac{h^{(2)}_{\mu\nu} h^{(2)}_{ t t,\nu}}{2} - \frac{h^{(4)}_{ t t,\mu}}{2} + h^{(3)}_{ t\mu, t} \bigg] \, .
\end{align}
These equations can be simplified even further by making use of the result $n_{A} = -n_{x} X_{,A}$.

Similarly, the leading-order parts of $tA$ and $AB$-components of the extrinsic curvature tensor are given by
\begin{align} \label{51}
{K}^{(1)}_{tA} &=  -X_{,A t} \, ,
\end{align}
and
\begin{align}
{K}^{(0)}_{AB} &=  -X_{,A B} \ .  \label{52}
\end{align}
These two equations, together with the result $K_{ij}=0$, imply that $X_{,A}$ is independent of $t$, $y$ and $z$ at lowest order. This implies that $X_{,A}$ also vanishes at lowest order, as $X_{,A}$ is forced by symmetry to vanish at the centre of every cell face. 

This information allows us to write simplified versions of the next-to-leading-order parts of the $tA$ and $AB$-components of the extrinsic curvature tensor as
\begin{align} \label{53}
{K}^{(3)}_{tA}&= -X^{(2)}_{,A t} + \frac{1}{2} \bigg[   h^{(3)}_{ tA,x} -  h^{(3)}_{ tx,A}  \bigg] - \frac{h^{(2)}_{ t t,A} X_{,t}}{2} \, ,
\end{align}
and
\begin{align} \label{54}
{K}^{(2)}_{AB} &=  -X^{(2)}_{,AB} +\frac{1}{2} n_{\alpha}h_{AB,\alpha}  \, .
\end{align}
This is all the information we require about the  extrinsic curvature of the boundaries in our lattices.

Using Eq. \eqref{23}, we can finally obtain from Eqs. \eqref{49} - \eqref{54} the conditions
\begin{align}  \label{55}
 X_{,tt} &= \bigg[ \Phi_{,x} - 2\Phi\Phi_{,x} + \frac{h^{(4)}_{tt,x}}{2} - h^{(3)}_{ tx, t} \\
 & \hspace{0.5cm} -3\Phi_{,x} X_{,t}^{2} -3 \Phi_{, t} X_{,t}  - X^{(2)}_{,A} \Phi_{,A} \bigg] \bigg|_{x=X} +  O(\epsilon^6) \,  ,  \nonumber 
 \end{align}
and
\begin{align} 
  X_{, tA} & = \frac{1}{2} \bigg[ h^{(3)}_{ tA,x} -  h^{(3)}_{ tx,A} - 4\Phi_{,A} X_{,t}\bigg] \bigg|_{x=X} +  O(\epsilon^5) \label{56} \, ,  
\end{align}
and
\begin{align}  
  X_{,AB} &= \delta_{AB}  \Phi_{,x} \Big|_{x=X} +  O(\epsilon^4) \ , \label{57}
\end{align}
where we have made explicit the requirement that each equation is to be evaluated on the boundary, at $x=X$. Eq. \eqref{55} is similar to the geodesic equation, as shown in Appendix \ref{AppendixA}.

\section{COSMOLOGICAL EXPANSION} \label{sec5}

We now have enough information to find the equations for the acceleration of the boundary of each cell, up to post-Newtonian accuracy. These will be the analogue of the Friedmann equations, of homogeneous and isotropic cosmological models. In this section, we begin by reproducing the lowest order Friedmann-like equations at Newtonian order. We then proceed to obtain the post-Newtonian contributions to the same equations.

\subsection{Newtonian Order} \label{sec5a}

We can begin by defining the gravitational mass within each cell as
\begin{align}  \label{58}
M &\equiv \int_{\Omega}{\rho} \ dV^{(0)} \ ,
\end{align}
where $dV^{(0)}$ is the spatial volume element at zeroth order, and $\Omega$ is now the spatial volume of the interior of a cell. We can apply Gauss' theorem to this equation, so that 
\begin{align}  \label{59}
4 \pi G M &= -\int_{\Omega}{\nabla^2  \Phi} \ dV^{(0)} \nonumber \\
&= -\int_{\partial \Omega} \bm{n} \cdot \nabla \Phi \ dA^{(0)} \ , 
\end{align}
where $\bm{n} = n_{\alpha}$ is the normal to a cell face ,and $dA$ is the area element of the boundary of the cell, $\partial \Omega$.  

By noting that the cell face is flat to lowest order (i.e. that $X_{,A}= O(\epsilon^2)$), we can see that it is possible to write $X=X(t)$. Together with the lowest-order part of Eq. \eqref{55}, this implies that $\bm{n} \cdot \nabla \Phi$ is constant on the boundary. We therefore have
\begin{align}  \label{60}
4 \pi G M &= -A \bm{n} \cdot \nabla \Phi \, ,
\end{align}
where $A$ is the total surface area of a cell. For a cell that is a regular polyhedron, we can take the surface area to be $A=\alpha_{\kappa} X^2$, where $X$ is the coordinate distance from the centre of the cell to the centre of the cell face, where $\kappa$ is the number of faces of the cell, and where $\alpha_{\kappa}$ are the numerical coefficients that are given in Table \ref{tab}.
  
The lowest-order part of Eq. \eqref{55}, along with Eq. \eqref{60}, then gives us
\begin{align}   \label{61}
X_{,tt} &= (\bm{n} \cdot \nabla \Phi)|_{x=X}=  -\frac{4\pi G M}{A} = -\frac{4\pi G M}{\alpha_{\kappa}X^2} \, .
\end{align}
We can solve this equation by multiplying both sides by $X_{,t}$. This gives
\begin{align}   \label{62}
\frac{1}{2} ((X_{,t})^2)_{,t}&=  -\frac{4\pi G M X_{,t}}{\alpha_{\kappa}X^2} \, ,
\end{align}
which can be integrated to find
\begin{align} \label{63} 
X_{,t} &=  \pm \sqrt{\frac{8\pi G M}{\alpha_{\kappa} X} - C} \, ,
\end{align}
where $C=C(y,z)$ is an integration constant in $t$. However, if this equation is to satisfy $X,_{ tA} = O(\epsilon^3)$, we see that $C$ must also be a constant in $y$ and $z$. Eq. \eqref{63} is similar in form to the Friedmann equation, with $X$ behaving like the scale factor, and the constant $C$ behaving like the Gaussian curvature of homogeneous spatial sections.

From now on, we will take the positive branch in Eq. \eqref{63}, which corresponds to an expanding universe. The solution to this equation depends on the sign of $C$. For $C= 0$ we have  
\begin{align} 
X &= \bigg(\frac{3}{2}\bigg) ^{2/3} \bigg(\sqrt{\frac{8\pi G M}{\alpha_{\kappa}}} t - t_{0}\bigg)^{2/3} +  O(\epsilon^2) \, ,  \label{65}
\end{align}
where $t_{0}$ is a constant, which can be absorbed into $t$ by a coordinate re-definition.

For $C\neq 0$, we can obtain parametric solutions. For $C> 0$ we have
\begin{align}   \label{66}
  X & = \frac{8\pi G M}{\alpha_{\kappa}|C|} \sin^2\bigg( \frac{\eta}{2}\bigg) \, , \nonumber \\ \nonumber \\
  t - t_{0} &= \frac{4\pi G M}{\alpha_{\kappa}|C|^{3/2}} (\eta - \sin \eta) \, ,
\end{align}
where $\eta = \int dt/X$ is the analogue of conformal time. Similarly, for $C< 0$, we get
\begin{align}  \label{67}
  X &= \frac{8\pi G M}{\alpha_{\kappa}C} \sinh^2\bigg( \frac{\eta}{2}\bigg) \ , \nonumber \\ \nonumber \\
  t - t_{0} &= \frac{4\pi G M}{\alpha_{\kappa}C^{3/2}} (\sinh \eta - \eta) \ .
\end{align}
These solutions represent parabolic, closed, and hyperbolic spaces, respectively. At this order, they expand in the same way as a dust-dominated homogeneous-and-isotropic geometry with the same total gravitational mass.

\subsection{Post-Newtonian Order} \label{sec5b}

To find the post-Newtonian contributions to the acceleration of the boundary of each cell it will be useful to replace all of the terms in Eq. \eqref{55} by more physically relevant quantities. The first step in doing this will be to include all orders in the calculation of Gauss' theorem, up to $ O(\epsilon^4)$. We can begin by expanding the normal derivative of the potential on the boundary to post-Newtonian order, so that
\begin{widetext}
\begin{align}  
 \quad \frac{1}{2}\int_{S} \bm{n} \cdot \nabla h_{tt} \ dA = \kappa \int_{S} \bigg(\Phi_{,x} +{n^{x}}^{(2)} \Phi_{,x} + {n^{A}}^{(2)} \Phi_{,A} + \frac{h^{(4)}_{tt,x}}{2}\bigg) \ dS^{(0)} + \kappa \int_{S} \Phi_{,x} dS^{(2)} +  O(\epsilon^6) \, ,  \label{68} 
\end{align}
\end{widetext}
where $dS$ is the area element of one face of the cell, where $\kappa$ is the number of faces of the cell, where $S$ is the area of one face of the cell, where $A = \kappa S$ is the total surface area of the cell, and where $dS^{(2)}$ is the $O(\epsilon^2)$ part of the area element.

We can then apply Gauss' theorem, making sure we use covariant derivatives, in order to ensure we include all post-Newtonian terms. This gives
\begin{align}
&\; \quad \frac{1}{2}\int_{S} \bm{n} \cdot \nabla h_{tt} \ dA \nonumber
\\&= \frac{1}{2} \int_{\Omega} {g^{(3)}}^{\mu\nu} h_{tt;\mu\nu} \ dV  \nonumber
\\&= \int_{\Omega} \bigg(\frac{1}{2}\nabla^2 h_{tt} -2\Phi\nabla^2\Phi + |\nabla \Phi|^2\bigg) \ dV +  O(\epsilon^6) \, .  \label{69}
\end{align}
Note that here $h_{tt}$ is being used to denote both $h^{(2)}_{tt}$ and $h^{(4)}_{tt}$. Expanding, and using the lower-order parts of the field equations, allows us to write the first and last terms on the right-hand side of Eq. \eqref{69} as
\begin{align}
\frac{1}{2}\int_{S} \nabla^2 h_{tt} \ dV &= -4\pi G \int_{\Omega} \rho \ dV^{(0)}  -4\pi G \int_{\Omega} \rho \ dV^{(2)} \nonumber \\&\quad + \frac{1}{2} \int_{\Omega}  \nabla^2 h^{(4)}_{tt} \ dV^{(0)} +  O(\epsilon^6) \, ,\label{70} 
\end{align}
and
\begin{align}
\int_{\Omega}{|\nabla \Phi|^2 } \ dV &=  \int_{\Omega} \bigg(4 \pi G \Phi \rho+ \frac{1}{2}\nabla^2 \Phi^2 \bigg) \ dV^{(0)}   \nonumber \\
&= 4\pi G\avg{\rho \Phi} + \kappa \int _{S} \Phi \Phi_{,x} \ dS^{(0)} \, ,  \label{71}
\end{align}
where $dV^{(2)}$ is the $O(\epsilon^2)$ correction to the volume element, and where we have introduced the new notation
\begin{align}
 \avg{\varphi} =  \int_{\Omega} \varphi \ dV \, ,  \label{72}
\end{align}
where $\varphi$ is some scalar function on the space-time.

To find the area and volume elements up to $ O(\epsilon^2)$ we need the induced 2-metric on the boundary, $g^{[2]}_{AB}$, and the spatial 3-metric, $g^{[3]}_{\mu\nu}$, both up to $O(\epsilon^2)$. These are given by 
\begin{align}  
g^{[2]}_{AB} &\equiv g^{[3]}_{AB}  - n_{A}n_{B}  \nonumber
%\\&= g^{[3]}_{AB} - n_{x}^2 X_{,A}X_{,B} \nonumber
\\&= (1+2\Phi) \delta_{AB}  +  O(\epsilon^4) \label{73} \, , 
\end{align}
and
\begin{align}
g^{[3]}_{\mu\nu} &=  (1+2\Phi) \delta_{\mu\nu}  +  O(\epsilon^4) \, .  \label{74}
\end{align}
The determinants of these two quantities, up to the required accuracy, are given by
\begin{align} \label{75}
det(g^{[2]}_{AB}) &= 1+ 4\Phi  +  O(\epsilon^4) \, , 
\end{align}
and
\begin{align}
det(g^{[3]}_{\mu\nu}) &= 1 + 6\Phi +  O(\epsilon^4) \, . \label{76}
\end{align} 
By taking the square root of these determinants, and Taylor expanding them, we obtain the higher-order area and volume elements in terms of their lower-order counterparts:
\begin{align}  \label{77}
dS^{(2)} &= 2\Phi \ dS^{(0)} \, , \\ 
dV^{(2)} &= 3\Phi \ dV^{(0)} \, . \label{78}
\end{align}
We can now evaluate the higher-order corrections to the normal. 

As $X_{,A}$ vanishes at lowest order, ${n^{A}}^{(2)}$ is given by  
\begin{align}  \label{79}
{n^{A}}^{(2)} = -X^{(2)}_{,A} \, .
\end{align}
We can also use the normalisation of the space-like normal, $n_{\alpha} n^{\alpha} =1$, to obtain
\begin{align}  \label{80}
 {n^{x}}^{(2)} = -\Phi \, . 
\end{align}
Using Eqs. \eqref{68}--\eqref{80} we can now write
\begin{align}  \label{81}
 &\kappa \int_{S} \bigg(\Phi_{,x} + \frac{ h^{(4)}_{tt,x}}{2}\bigg) \ dS \nonumber
 \\ =& -4\pi G M  + \frac{1}{2} \int_{\Omega}  \nabla^2 h^{(4)}_{tt} \ dV^{(0)} + 2\kappa \int _{S} \Phi \Phi_{,x} \ dS^{(0)} \nonumber
 \\& + \kappa \int_{S}  X^{(2)}_{,A} \Phi_{,A}  \ dS^{(0)} \ . 
\end{align}
To understand this equation further, we note that the second term on the right-hand side can be written as
\begin{widetext}
\begin{align} 
\frac{1}{2} \int_{\Omega}  \nabla^2 h^{(4)}_{tt} \ dV^{(0)} &=  \int_{\Omega} (-\nabla^2 \Phi^2 + 2\nabla^2\Phi_{1} + 2\nabla^2 \Phi_{2} + \nabla^2 \Phi_{3} + 3\nabla^2 \Phi_{4}) \ dV^{(0)} \nonumber  \\
& = -\int_{S} \bm{n} \cdot \nabla \Phi^2 \ dA^{(0)} + \int_{\Omega} (2\nabla^2\Phi_{1} + 2\nabla^2 \Phi_{2} + \nabla^2 \Phi_{3} + 3\nabla^2 \Phi_{4}) \ dV^{(0)} \nonumber  \\
& = - 2 \kappa \int_{S} \Phi \Phi_{,x}\ dS^{(0)}   + \int_{\Omega} (2\nabla^2\Phi_{1} + 2\nabla^2 \Phi_{2} + \nabla^2 \Phi_{3} + 3\nabla^2 \Phi_{4}) \ dV^{(0)}  \ , \label{82}
% X_{,tt}^{(2)} &=  n_{x}[ - 4\Phi \Phi_{,x}  + 2\Phi_{1,x} + 2 \Phi_{2 ,x} +  \Phi_{3 ,x} + 3 \Phi_{4,x} - h_{ tx, t} -3\Phi_{,x} X_{,t}^{2}-3 \Phi_{, t} X_{,t}] \nonumber 
\end{align} 
\end{widetext}
where we have now used Eq. \eqref{44} and Gauss' theorem. Using Eqs. \eqref{72} and \eqref{82}, we can now re-write Eq. \eqref{81} as
\begin{align}  
&\kappa \int_{S} \bigg[\Phi_{,x}  + \frac{h^{(4)}_{tt,x}}{2}  -  X^{(2)}_{,A} \Phi_{,A}\bigg] \ dS \nonumber
\\=&-4\pi G M - 8\pi G\avg{\rho v^2} - 8\pi G\avg{\rho \Phi}  \nonumber
\\&-4\pi G \avg{\rho \Pi} - 12\pi G\avg{p} \, .\label{83}
\end{align}

To proceed further, let us now determine the functional form of $X$ up to $O(\epsilon^2)$. Using the lowest order parts of Eqs. \eqref{55} and \eqref{57}, this is given by
\begin{align} \label{84}
 X = \zeta(t) + \frac{1}{2} (y^2+ z^2) \bm{n} \cdot \nabla \Phi + O(\epsilon^4) \ ,
\end{align}  
where $\zeta(t)$ is some function of time only. Taking time derivatives, and substituting from Eq. \eqref{55}, then gives 
\begin{align} 
\zeta_{,tt} =& X_{,tt}  - \frac{1}{2}(y^2+ z^2)(\bm{n} \cdot \nabla \Phi)^{..} + O(\epsilon^6) \ \nonumber
\\ =& \Phi_{,x}- 2\Phi\Phi_{,x} + \frac{h^{(4)}_{tt,x}}{2} - h^{(3)}_{ tx, t} -3\Phi_{,x} X_{,t}^{2} -3 \Phi_{, t} X_{,t}  \nonumber
\\&- X^{(2)}_{,A} \Phi_{,A} - \frac{1}{2}(y^2+ z^2)(\bm{n} \cdot \nabla \Phi)^{..}+ O(\epsilon^6) \, ,  \label{85}
\end{align} 
where all terms in this equation should be taken as being evaluated on the boundary. 
 
Many of the terms in this equation can be simplified using the lower-order solutions. For example, using Eqs. \eqref{61} and \eqref{63}, and taking time derivatives, gives
%\begin{align} 
%\bm{n} \cdot \nabla \Phi &= \frac{\partial \Phi}{\partial X}|_{X} = -\frac{4\pi G M}{\alpha_{\kappa}X^2} \ , \nonumber \\ \nonumber \\
%\bm{n} \cdot \nabla \dot{\Phi} &= \frac{8\pi G M}{\alpha_{\kappa}X^3} X_{,t} \ , \nonumber \\ \nonumber \\ 
%\bm{n} \cdot \nabla\ddot{\Phi} & =  -\frac{24\pi G M}{\alpha_{\kappa}X^4} X_{,t}^2 +  \frac{8\pi G M}{\alpha_{\kappa}X^3} X_{,tt} \nonumber \\ \nonumber \\
% &=  -\frac{192\pi^2 G^2 M^2}{\alpha_{\kappa}^2 X^5} -\frac{32\pi^2 G^2 M^2}{\alpha_{\kappa}^2 X^5}  + \frac{24\pi G M C}{\alpha_{\kappa}X^4}   \ .     \label{91}
%\end{align}
\begin{align} 
(\bm{n} \cdot \nabla \Phi)^{..} & = -\frac{224\pi^2 G^2 M^2}{\alpha_{\kappa}^2 X^5}  +\frac{24\pi G M C}{\alpha_{\kappa}X^4}   \, .     \label{86}
\end{align}
We are now in a position to express the equation of motion in terms of variables that can be easily associated with the matter fields in the space-time. Recall that the total surface area of the cell at lowest order is given by $A = \kappa S = \alpha_{\kappa} (X^{(0)})^2$, where $X^{(0)}$ is the zeroth-order part of $X$, which we solved for earlier. As Eq. \eqref{85} is a function of $t$ only, we can integrate over the area on a cell face to obtain
%
%\begin{align}   \label{87}
%  \int_{S} \zeta_{,tt} \ dA &=  \kappa \int_{S} \bigg[\Phi_{,x} - 2\Phi\Phi_{,x} + %\frac{h^{(4)}_{tt,x}}{2} - h_{ tx, t}   -3\Phi_{,x} X_{,t}^{2} -3 \Phi_{, t} X_{,t}   - %X^{(2)}_{,A} \Phi_{,A} - \frac{1}{2}  (y^2+ z^2) \bm{n} \cdot \nabla  \ddot{\Phi} \bigg]  \ dS %+ O(\epsilon^6) \ .  
% \end{align} 
% Evaluating the integral on the left-hand side, and substituting in Eqs. \eqref{83} and %\eqref{86}, gives
%\begin{widetext}
%\begin{align}  \label{88}
%A \zeta_{,tt}  &=  \kappa \int_{S} \bigg[\Phi_{,x} - 3\Phi\Phi_{,x} - \frac{1}{2} X_{,t}^2 %\Phi_{,x} - h_{ tx, t}  -3\Phi_{,x} X_{,t}^{2} -3 \Phi_{, t} X_{,t}  + \bigg(\frac{112\pi^2 %G^2 M^2}{\alpha_{\kappa}^2 X^5}  -\frac{12\pi G M C}{\alpha_{\kappa}X^4} \bigg)(y^2+ %z^2)\bigg] \ dS \nonumber \\ \nonumber \\
%&\quad  - 8\pi G \avg{\rho v^2} - 4\pi G \avg{\rho \Phi} -4\pi G \avg{\rho \Pi} - 12\pi G %\avg{p} + O(\epsilon^6) \, , 
%\end{align} 
\begin{align}  
&A \zeta_{,tt}  \nonumber
\\=&  -  4\pi G M +  \frac{\kappa S}{\alpha_{\kappa}X^2}\bigg(  \frac{96\pi^2 G^2 M^2}{\alpha_{\kappa}X} - 12\pi G M C\bigg) \nonumber
\\ &+ \kappa \int_{S} \bigg(\frac{8 \pi G M\Phi}{\alpha_{\kappa}X^2} - h_{ tx, t}  -3 \Phi_{, t} X_{,t} \bigg) \ dS   \nonumber \\
&- 8\pi G \avg{\rho v^2}- 8\pi G \avg{\rho \Phi} -4\pi G \avg{\rho \Pi} - 12\pi G \avg{p}  \nonumber
\\&+ \kappa \bigg(\frac{112\pi^2 G^2 M^2}{\alpha_{\kappa}^2 X^5}  -\frac{12\pi G M C}{\alpha_{\kappa}X^4} \bigg) \int_{S}(y^2+ z^2) \ dS + O(\epsilon^6) \, , \label{89}
\end{align}
where we have used Eqs. \eqref{83} and \eqref{86}, and substituted in for lower-order solutions.

We can make use of our gauge condition, $h_{ t\nu,\nu}= \frac{1}{2} h_{\nu\nu, t} = 3\Phi_{, t}$, and Gauss' theorem, to replace one of the terms in this equation in the following way:
\begin{align}  \label{90}
 \kappa \int_{S} n_{\alpha}h_{ t\alpha, t} \ dS &=  3 \int_{\Omega} \Phi_{, t t} \ dV \ .
\end{align}
%We must also expand our limits of integration to $O(\epsilon^2)$, when dealing with the lowest-order term, in order to ensure we include all the required post-Newtonian terms. This gives
%\begin{align}
% - \frac{4\pi G M}{\alpha_{\kappa} X^2} 
%%&=  - \frac{4\pi G M}{\alpha_{\kappa} (X^{(0)})^2}\bigg(1 + \frac{X^{(2)}}{X^{(0)}}\bigg)^{-2} %\nonumber  \\
%&= - \frac{4\pi G M}{\alpha_{\kappa} (X^{(0)})^2} + \frac{8\pi G M X^{(2)}}{\alpha_{\kappa} (X^{(0)})^3} \ , \label{91}
% \end{align}
%where $X^{(0)}$ is the zeroth-order part of $X$, which we solved for earlier. 
Using Eqs. \eqref{90}, we can write Eq. \eqref{89} as
\begin{align} \label{92}
&A \zeta_{,tt}  \nonumber
\\=&  -  4\pi G M  + \frac{\kappa S}{\alpha_{\kappa}(X^{(0)})^2}\bigg( \frac{96\pi^2 G^2 M^2}{\alpha_{\kappa}X^{(0)}}- 12\pi G M C \bigg) \nonumber
\\&+ \kappa  \int_{S} \bigg(\frac{8 \pi G M \Phi}{\alpha_{\kappa}(X^{(0)})^2} -3 \Phi_{, t} X^{(0)}_{,t} \bigg) \ dS  - 3 \int_{\Omega}  \Phi_{, t t} \ dV  \nonumber
\\&- 8\pi G \avg{\rho v^2} - 8\pi G \avg{\rho \Phi} -4\pi G \avg{\rho \Pi} - 12\pi G \avg{p}  
\nonumber  
\\&+ \kappa \bigg(\frac{112\pi^2 G^2 M^2}{\alpha_{\kappa}^2 (X^{(0)})^5}  -\frac{12\pi G M C}{\alpha_{\kappa}(X^{(0)})^4} \bigg)\int_{S}(y^2+ z^2) \ dS + O(\epsilon^6)  \, .  
\end{align} 
Then we must find the $O(\epsilon^2)$ correction to the surface area. This is to ensure that we include all $O(\epsilon^4)$ corrections to the equation of motion. To find the $O(\epsilon^2)$ correction to the surface area, we must take into account that the edges of the boundaries are curved at $O(\epsilon^2)$ and the background is also curved at $O(\epsilon^2)$. Then the surface area is given by
\begin{align}
A = \kappa S = \kappa \int_{S} (1 + 2\Phi) \ dS + O(\epsilon^4)
\end{align}
In general, this is dependent on the cell shape.
Therefore, the equation of motion of the boundaries can be written in its final form as
\begin{widetext}
\begin{align} \label{93}
 X_{,tt} =& -  \frac{4\pi G M}{A} -\frac{12\pi G M C}{\alpha_{\kappa}(X^{(0)})^2}  -\frac{4\pi G}{\alpha_{\kappa}(X^{(0)})^2} \bigg[  2\avg{\rho v^2} + 2\avg{\rho \Phi} + \avg{\rho \Pi} +  3\avg{p} \bigg] \nonumber \\ &+\frac{\kappa}{\alpha_{\kappa}(X^{(0)})^2} \int_{S} \bigg(\frac{8 \pi G M \Phi}{\alpha_{\kappa}(X^{(0)})^2}-3 \Phi_{, t} X^{(0)}_{,t} \bigg) \ dS     - \frac{3}{\alpha_{\kappa}(X^{(0)})^2} \int_{\Omega}  \Phi_{, t t} \ dV  + \frac{1}{\alpha_{\kappa}(X^{(0)})^3} \bigg[\frac{96\pi^2 G^2 M^2}{\alpha_{\kappa}}\bigg]   \nonumber
\\ &+  \bigg(\frac{112\pi^2 G^2 M^2}{\alpha_{\kappa}^2 (X^{(0)})^5}  - \frac{12\pi G M C}{\alpha_{\kappa}(X^{(0)})^4} \bigg)\bigg[\frac{\kappa}{\alpha_{\kappa}(X^{(0)})^2} \int_{S}(y^2+ z^2) \ dS - (y^2 +z^2)\bigg]+ O(\epsilon^6)  \, . 
\end{align}
\end{widetext}
This gives us the post-Newtonian correction to the acceleration of the boundary, and hence the post-Newtonian correction to the accelerated expansion of the Universe. This equation is the main result of this section.

The first term in Eq. \eqref{93} is the standard Friedmann-like term for a dust-like source. The second term is a higher-order correction due to the presence of the spatial curvature-like term. The third term in the first line contains all post-Newtonian corrections to the matter sector. In the second line, the terms integrated over the area and volume are dependent on the potential, $\Phi$, and the rate of change of the potential. The terms that go as $X^{-3}$ behave like radiation terms, when compared to the standard Friedmann equation. However, we remind the reader that these terms are purely a result of geometry, and the non-linearity of Einstein's equations. In the last line of this equation, the post-Newtonian contributions depend on the cell shape that is being considered.

\section{POST-NEWTONIAN COSMOLOGICAL SOLUTIONS} \label{sec6}

In this section we will present solutions to Eq. \eqref{93}. We start by specialising this equation to lattices constructed from cubic cells. In this case we have $\alpha_{\kappa} = 24$, and $\kappa=6$. The total surface area is given by
\begin{align}
%A= 6( 4\zeta^2 - \frac{8}{9 X^2} \pi G M X^3) + 12\int_{S} \Phi \ dS =  24\zeta^2 - \frac{16 \pi G M X}{3} + 12\int_{S} \Phi \ dS + O(\epsilon^4)
A= 24\zeta^2 - \frac{16 \pi G M X}{3} + 12\int_{S} \Phi \ dS + O(\epsilon^4)
\end{align} 
where $\zeta$ is the time-dependent part of $X$. The lowest-order part of the limits of integration in both the $y$ and $z$-directions are also given simply by $-X^{(0)}$ and $X^{(0)}$. The equation of motion, given in Eq. \eqref{93}, then simplifies to
\begin{align} \nonumber
 &X_{,tt} \\ \hspace{-0.5cm} =& -\frac{\pi G}{6\zeta^2} \bigg[  M + 5 M C + 2  \avg{\rho v^2} + 2 \avg{\rho \Phi} + \avg{\rho \Pi} + 3 \avg{p}\bigg]   \nonumber \\ &+  \frac{1}{8(X^{(0)})^2} \bigg[ \int_{S} \bigg(\frac{4\Phi \pi G M}{3(X^{(0)})^2} -6 \Phi_{, t} X^{(0)}_{,t} \bigg) \ dS   - \int_{\Omega}  \Phi_{, t t} \ dV \bigg] \nonumber
 \\&+ \frac{1}{(X^{(0)})^3} \bigg[\frac{7\pi^2 G^2 M^2}{27} \bigg]   \nonumber  \\
&- \bigg(\frac{7\pi^2 G^2 M^2}{36 (X^{(0)})^5}  -\frac{\pi G M C}{2(X^{(0)})^4} \bigg)(y^2 +z^2)+ O(\epsilon^6)  \, . \label{94}
\end{align}
The last term in this equation is a function of its position on the boundary, and vanishes at the centre of a cell face.

We can solve Eq. \eqref{94} if we know the functional form of the potential, $\Phi$, and its time dependence, as well as that of the post-Newtonian corrections to the matter sector. However, we do not need to know the functional form of all the post-Newtonian corrections, as we can replace one of the higher-order terms with the conserved post-Newtonian mass. This is given by
\begin{equation} \label{95}
M_{PN} = \int_{V} \rho \bigg(\frac{1}{2} v^2 + 3 \Phi \bigg)\ dV = \frac{1}{2}\avg{\rho v^2} + 3\avg{ \rho\Phi} \, . 
\end{equation}
The proof that this object is conserved can be found in Appendix \ref{AppendixB}. 

We can now find the general functional form of the potential $\Phi$ for our model, using the Green's function formalism. We will do this below for the case of cubic cells. Similar analyses can also be performed for the other platonic solids. This result can then be used to evaluate the acceleration of the boundary. After this, we proceed to study the special case of point sources, where the form of the potential $\Phi$ can be found somewhat more straightforwardly, and where the first post-Newtonian correction to the acceleration of the boundary can be determined explicitly.

\subsection{The General Solution: An Application of the Green's Function Formalism} \label{sec6a}

\begin{figure}[t!]
    \includegraphics[width=85mm]{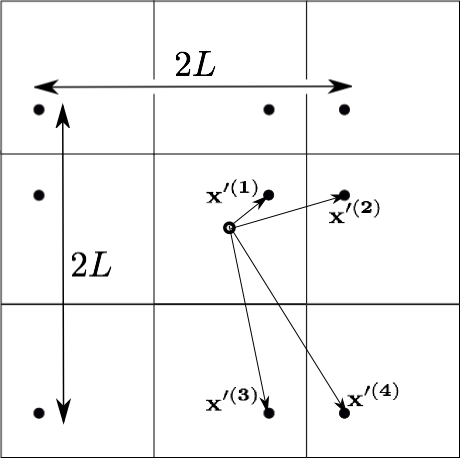}
      \caption{\label{fig3} A 2-D representation of the vectors used to locate the position of image points. In two dimensions we require only four unique vectors, as compared to eight in 3-D. The four lattice vectors are given by $\mathbf{x'^{(1)}} = \mathbf{x'}$, $\mathbf{x'^{(2)}} = \mathbf{x'} + L\mathbf{e_{1}}- 2x_{1}'\mathbf{ e_{1}}$, $\mathbf{x'^{(3)}} = \mathbf{x'} + L\mathbf{e_{2}}- 2x_{2}' \mathbf{e_{2}}$, and $\mathbf{x'^{(4)}} = -\mathbf{x'} + L(\mathbf{e_{1} + e_{2}})$.}
   \end{figure}

In this section we will use the explanation of Green's functions from Section \ref{sec2c}, and in particular Eq. \eqref{102}. To give a concrete example of how this works, let us now consider a lattice of cubes arranged on $\mathbb{R}^3$. In Figure \ref{fig3} we show a 2-D representation of such a 3-D lattice. As before, we assume reflection symmetry about every boundary, which imposes a periodicity on our structure. For cubic cells of edge length $L= 2X$, the periodicity of the the lattice will be $2L$. That is, if we move a distance of $2L$ in any direction in our lattice, then two reflections will ensure we return to a point that is identical to our starting position. 

If we consider one example cell, then we can now use the method of images to construct a Green's function that is symmetric around each of its boundaries, and that therefore satisfies the required boundary condition at each of its faces, $\bm{n} \cdot \nabla \mathcal{G} |_{\partial \Omega} = 0$. Due to the identical nature of every cell, such a Green's function can then be re-used for each of the cells. The way that this method will work is by introducing mirror images of the points in our original cell. We therefore consider the point sources of the Green's function to be a set of Dirac delta functions, separated from infinitely many identical point sources by pairwise distances of $2L$. The structure that results will be a superposition of several `Dirac combs'.  

A Dirac comb can be expressed as a Fourier series in the following way:
\begin{align}  \label{104}
\sum_{\bm{\beta}\in \mathbb{Z}^3}\delta(\mathbf{x} - 2L\bm{\beta}) = \sum_{\bm{\beta}\in \mathbb{Z}^3} \frac{1}{8L^3} e^{\pi i\bm{\beta} \cdot \frac{\mathbf{x}}{L}} \ , 
\end{align} 
where $\bm{\beta}= (\beta_{1}, \beta_{2}, \beta_{3})$, and where $\beta_{1}$, $\beta_{2}$ and $\beta_{3}$ are integers. To construct our Green's function, we must include the location of image points, in relation to the location of points in the central cell. In Figure \ref{fig3}, in 2-D, we choose an arbitrary point in the example cell, and use $\mathbf{x^{(1)}}$ to represent its position with respect to the centre of that cell. The mirror symmetry across the boundary of every cell results in $8$ image points in the $8$ surrounding cells. However, we only require $4$ unique vectors to describe the positions of the initial point source and its images. These $4$ vectors are shown in Figure \ref{fig3}. The other points can be added by considering Dirac combs, with periodicity $2 L$, that contain these initial $4$ points.
 
In 3-D we can do something similar, but we require $8$ unique vectors to describe the position of the initial point and its images. The source for the Green's function is then the sum of the Dirac combs that contain all $8$ of the image points described by these $8$ position vectors, plus an additive time dependent constant. Using Eqs. \eqref{98} and \eqref{104}, the source function is thus given by
\begin{align} \label{105}
\nabla^2 \mathcal{G} &= -\sum_{j=1}^{8}\sum_{\bm{\beta}\in \mathbb{Z}^3} \frac{1}{8L^3} e^{\pi i\bm{\beta} \cdot \frac{(\mathbf{x-x'^{(j)}})}{L}} + \frac{1}{L^3} \\ \nonumber \\
&= -\sum_{j=1}^{8}\sum_{\bm{\beta}\in \mathbb{Z}_{*}^3} \frac{1}{8L^3} e^{\pi i\bm{\beta} \cdot \frac{(\mathbf{x-x'^{(j)}})}{L}} \ , \label{106}
 \end{align} 
where $\mathbb{Z}_{*}^3$ does not include the null triplet $\bm{\beta}=(0,0,0)$, and where the $\mathbf{x'^{(j)}}$ are given by
\begin{align} \label{107}
\mathbf{x'^{(1)}} &= \mathbf{x'} \ ,  \nonumber \\  \nonumber \\
\mathbf{x'^{(2)}} &= \mathbf{x'} + L\mathbf{e_{1}}- 2x_{1}' \mathbf{e_{1}} \ ,  \nonumber \\  \nonumber \\
\mathbf{x'^{(3)}} &= \mathbf{x'} + L\mathbf{e_{2}}- 2x_{2}' \mathbf{e_{2}} \ ,  \nonumber \\  \nonumber \\
\mathbf{x'^{(4)}} &= -\mathbf{x'} + L(\mathbf{e_{1} + e_{2}})+ 2x_{3}'\mathbf{e_{3}} \ ,  \nonumber \\  \nonumber \\
\mathbf{x'^{(5)}} &= \mathbf{x'} + L\mathbf{e_{3}}- 2x_{3}' \mathbf{e_{3}} \ ,  \nonumber \\  \nonumber \\
\mathbf{x'^{(6)}} &= -\mathbf{x'} + L(\mathbf{e_{1} + e_{3}})+ 2x_{2}'\mathbf{ e_{2}} \ ,  \nonumber \\  \nonumber \\
\mathbf{x'^{(7)}} &= -\mathbf{x'} + L(\mathbf{e_{2} + e_{3}})+ 2x_{1}'\mathbf{ e_{1}} \ ,  \nonumber \\  \nonumber \\
\mathbf{x'^{(8)}} &= -\mathbf{x'} + L(\mathbf{e_{1} + e_{2}+ e_{3}}) \ , 
\end{align} 
where $x_{1}' = \mathbf{x'\cdot e_{1}}$, $x_{2}' =\mathbf{ x' \cdot e_{2}}$ and $x_{3}' =\mathbf{ x'\cdot e_{3}}$, and where $\mathbf{e_{1}}$, $\mathbf{e_{2}}$ and $\mathbf{e_{3}}$ are orthogonal unit vectors. 

The solution to Eq. \eqref{106} is then given by
\begin{align} \label{108}
\mathcal{G}(\mathbf{x}, \mathbf{x'}, t)= \sum_{j=1}^{8}\sum_{\bm{\beta}\in \mathbb{Z}_{*}^3} \frac{1}{8\pi^2L|\bm{\beta}|^2} e^{\pi i\bm{\beta} \cdot \frac{(\mathbf{x-x'^{(j)}})}{L}} \, .
\end{align} 
If we take, as an example equation, the Newton-Poisson equation \eqref{13}, we can then see from Eq. \eqref{102} that the Newtonian potential $\Phi$ is given by
\begin{align} \label{109}
\Phi &= \bar{\Phi}+ 4\pi G \int_{\Omega} \mathcal{G} \rho \, dV - \frac{\pi G M}{6} \int_{\partial \Omega} \frac{\mathcal{G}}{X^2} \, dA \, , 
\end{align} 
where we have used Eq. \eqref{61}. This is the general solution for the potential. That is, for  any given energy density distribution we can simply evaluate the integral to find the potential, as well as the rate of change of the potential. In Eq. \eqref{94}, this, along with the mass, velocity and pressure of the matter fields can be used to evaluate the acceleration of the boundary numerically, up to post-Newtonian accuracy. Hence, this expression allows us to find the post-Newtonian correction to the expansion of the Universe for general configurations of matter.
 
Of course, within the interior of each cell we can also solve for each of the post-Newtonian potentials defined in Section \ref{sec2b} using the same Green's function, as each of these potentials is defined as the solution to a Poisson equation. We therefore have a complete solution, for the motion of the boundary of every cell, and for the geometry interior to each cell.

\subsection{A Special Case: Point Sources} \label{sec6b}

In order to find an even more explicit solution, let us now consider the case of a single point mass, located at the centre of each cell. In this case, the Poisson equation simplifies to 
\begin{align} \label{110}
\nabla^2 \Phi = -4\pi G M \delta(\mathbf{x}) \ , 
\end{align}
where $M$ is the mass we defined in Eq. \eqref{58}. We can again use the method of images to solve for $\Phi$. In this case we want to place our image points such that $\Phi$ satisfies the inhomogeneous boundary condition given in Eq. \eqref{61}. We therefore place an image mass at the centre of each surrounding cell, so that image masses are separated by a distance $L$ from each other. We can then express the source of the potential, $\Phi$, as a sum of Dirac delta functions that correspond to these masses. We then continue by placing image masses at the centre of every cell that surrounds the cells that already contain image masses (taking care not to place two masses in any one given cell). We repeat this process $\mathcal{N}$ times, and then let $\mathcal{N} \rightarrow \infty$.

This description may initially sound similar to the process used to find the Green's function, above. There is, however, a subtle difference. In order for there to be a non-zero normal derivative of $\Phi$ on the boundary, we need to take a sum whose number of terms tends to infinity, rather than an array that is infinitely extended from the outset. These two things are not equivalent, in this case.

The source of the potential can then be written as
\begin{align}  \label{111}
\nabla^2 \Phi = -4\pi G M  \lim _{\mathcal{N} \to \infty} \sum_{\bm{\beta} = - \mathcal{N}}^{\mathcal{N}} \delta(\mathbf{x} - L \bm{\beta}) \ , 
\end{align}
where $L= 2X$ is again the edge length of the cubic cell, and where $\mathcal{N}$ is a positive integer. The solution to Eq. \eqref{111} is given by
\begin{align} \label{112}
\Phi &=  \lim _{\mathcal{N} \to \infty} \sum_{\bm{\beta} = - \mathcal{N}}^{\mathcal{N}}\frac{G M}{|\mathbf{x} - L \bm{\beta}|} + f(t) \, ,%\nonumber 
% \\ %&=  \lim _{\mathcal{N} \to \infty} \sum_{\bm{\beta} = - \mathcal{N}}^{\mathcal{N}}\frac{G M}{\sqrt{(x -\beta_{1}L)^2 + (y -\beta_{2}L)^2 + (z-\beta_{3}L)^2 }} + f(t) \ , 
\end{align}
where $f(t)$ is an arbitrary function of time. 

We can use $f(t)$ to regularize the value of $\Phi$ at $\mathbf{x} = 0$, such that it reduces to the regular form for a Newtonian potential around a single point source. This can be done by subtracting the contribution of all image points to the potential at $\mathbf{x} = 0$:
\begin{align} \label{113}
\Phi &=   \lim _{\mathcal{N} \to \infty} \sum_{\bm{\beta} = - \mathcal{N}}^{\mathcal{N}}\frac{G M}{|\mathbf{x} - L \bm{\beta}|} -  \lim _{\mathcal{N} \to \infty} \sum_{\bm{\beta^{*}} = - \mathcal{N}}^{\mathcal{N}}\frac{G M}{|L \bm{\beta}|} \, , 
%\nonumber \\ \nonumber \\ 
%&=  \lim _{\mathcal{N} \to \infty} \sum_{\bm{\beta} = - \mathcal{N}}^{\mathcal{N}}\frac{G M}{\sqrt{(x -\beta_{1}L)^2 + (y -\beta_{2}L)^2 + (z-\beta_{3}L)^2 }} -   \lim _{\mathcal{N} \to \infty} \sum_{\bm{\beta^{*}} = - \mathcal{N}}^{\mathcal{N}}\frac{G M}{\sqrt{\beta_{1}^2 + \beta_{2}^2 + \beta_{3}^2 }L} \ , 
\end{align}
where $\bm{\beta^{*}}$ indicates that the null triplet $\bm{\beta} = (0,0,0)$ is excluded from the sum. With this choice, the value of $\Phi$ near $\mathbf{x} = 0$ does not change as the number of image masses is increased. This can be considered as a boundary condition imposed at the location of the mass.

As $L$ is the only time dependent quantity in Eq. \eqref{113}, it can be seen that the rate of change of $\Phi$ is simply given by
\begin{align} 
\hspace{-0.2cm}\Phi_{, t} =&  \lim _{\mathcal{N} \to \infty} \sum_{\bm{\beta} = - \mathcal{N}}^{\mathcal{N}}\frac{G M( \bm{\beta} \cdot \mathbf{x} - |\bm{\beta}|^2 L) L_{,t}}{|\mathbf{x} - L \bm{\beta}|^{3}} 
\nonumber\\&+  
\lim _{\mathcal{N} \to \infty} \sum_{\bm{\beta^{*}} = - \mathcal{N}}^{\mathcal{N}}\frac{G ML_{,t}}{|\bm{\beta}|L^2} \ . \label{114}
\end{align}
Likewise, the second time derivative of $\Phi$ is given by
\begin{align} 
\Phi_{, t t} =&   \lim _{\mathcal{N} \to \infty} \sum_{\bm{\beta} = - \mathcal{N}}^{\mathcal{N}}\frac{G M( \bm{\beta} \cdot \mathbf{x} - |\bm{\beta}|^2 L) L_{,tt} - |\bm{\beta}|^2 G M{L_{,t}}^2}{|\mathbf{x} - \bm{\beta}L|^3}
 \nonumber \\
&+ \lim _{\mathcal{N} \to \infty} \sum_{\bm{\beta} = - \mathcal{N}}^{\mathcal{N}}\frac{3G M( \bm{\beta} \cdot \mathbf{x}- |\bm{\beta}|^2 L)^2 L_{,t}^2 }{|\mathbf{x} - \bm{\beta}L|^5} \nonumber \\ 
&+  \lim _{\mathcal{N} \to \infty} \sum_{\bm{\beta^{*}} = - \mathcal{N}}^{\mathcal{N}}\frac{G ML_{,tt}}{|\bm{\beta}|L^2} -   \lim _{\mathcal{N} \to \infty} \sum_{\bm{\beta^{*}} = - \mathcal{N}}^{\mathcal{N}}\frac{2G M{L_{,t}}^2}{|\bm{\beta}|L^3} \ . \nonumber  %\label{115}
\end{align}

Now, in order to solve Eq. \eqref{94}, we need to evaluate a few integrals. We need $\Phi$ and $\Phi_{, t}$ integrated over the boundary of a cell, and $\Phi_{, t t}$ integrated over the volume. These integrals are given explicitly below. Firstly,
\begin{widetext}
\begin{align}
\int_{-L/2}^{L/2} \Phi|_{x=L/2} \ dy dz =  \int_{-L/2}^{L/2} \bigg[ \lim _{\mathcal{N} \to \infty} \sum_{\bm{\beta} = - \mathcal{N}}^{\mathcal{N}}\frac{G M}{\sqrt{(L/2 -\beta_{1}L)^2 + (y -\beta_{2}L)^2 + (z-\beta_{3}L)^2 }} -   \lim _{\mathcal{N} \to \infty} \sum_{\bm{\beta^{*}} = - \mathcal{N}}^{\mathcal{N}}\frac{G M}{|\bm{\beta}|L}\bigg] \ dy dz \ .  \nonumber 
\end{align}
This can be simplified by redefining coordinates such that $y = L\hat{y}$, and $z = L \hat{z}$. This gives
\begin{align}  
%\int_{-L/2}^{L/2} \Phi|_{x=L/2} \ dy dz &= \frac{1}{L} \bigg[ \int_{-1/2}^{1/2} \bigg[ \lim _{\mathcal{N} \to \infty} \sum_{\bm{\beta} = - \mathcal{N}}^{\mathcal{N}}\frac{G M}{\sqrt{(1/2 -\beta_{1})^2 + (\hat{y} -\beta_{2})^2 + (\hat{z}-\beta_{3})^2 }} -   \lim _{\mathcal{N} \to \infty} \sum_{\bm{\beta^{*}} = - \mathcal{N}}^{\mathcal{N}}\frac{G M}{|\bm{\beta}|}\bigg] \quad  L^2 d\hat{y} d\hat{z} \bigg]  \nonumber\\
\int_{-L/2}^{L/2} \Phi|_{x=L/2} \ dy dz &=  G M L \int_{-1/2}^{1/2} \bigg[ \lim _{\mathcal{N} \to \infty} \sum_{\bm{\beta} = - \mathcal{N}}^{\mathcal{N}}\frac{1}{\sqrt{(1/2 -\beta_{1})^2 + (\hat{y} -\beta_{2})^2 + (\hat{z}-\beta_{3})^2 }}  -   \lim _{\mathcal{N} \to \infty} \sum_{\bm{\beta^{*}} = - \mathcal{N}}^{\mathcal{N}}\frac{1}{|\bm{\beta}|}\bigg] \ d\hat{y} d\hat{z}    \nonumber\\
 &\equiv  G M DL \ , \label{116}  %=  2 G M DX^{(0)} 
\end{align}
where the last line of this equation defines the quantity $D$. Secondly, $\Phi_{,t}$ integrated over the boundary is given by
\begin{align}  
\int_{-L/2}^{L/2} \Phi_{, t}|_{x=L/2} \ dy dz =  \int_{-L/2}^{L/2} \bigg[ \lim _{\mathcal{N} \to \infty} \sum_{\bm{\beta} = - \mathcal{N}}^{\mathcal{N}}\frac{G M( \beta_{1}x+\beta_{2}y+\beta_{3}z - |\bm{\beta}|^2 L) L_{,t}}{[(x -\beta_{1}L)^2 + (y -\beta_{2}L)^2 + (z-\beta_{3}L)^2 ]^{3/2}} +   \lim _{\mathcal{N} \to \infty} \sum_{\bm{\beta^{*}} = - \mathcal{N}}^{\mathcal{N}}\frac{G ML_{,t}}{|\bm{\beta}|L^2} \bigg] \ dy dz \ . \nonumber
\end{align}
We can again simplify this integral by using $\hat{y}$ and $\hat{z}$ coordinates. This gives
\begin{align} 
%\int_{-L/2}^{L/2} \Phi_{, t}|_{x=L/2} \ dy dz &=  \int_{-1/2}^{1/2} \bigg[ \lim _{\mathcal{N} \to \infty} \sum_{\bm{\beta} = - \mathcal{N}}^{\mathcal{N}}\frac{ G ML( \beta_{1}(1/2)+\beta_{2}\hat{y}+\beta_{3}\hat{z} - |\bm{\beta}|^2) L_{,t}}{L^3[(1/2 -\beta_{1})^2 + (\hat{y} -\beta_{2})^2 + (\hat{z}-\beta_{3})^2]^{3/2}} +   \lim _{\mathcal{N} \to \infty} \sum_{\bm{\beta^{*}} = - \mathcal{N}}^{\mathcal{N}}\frac{G ML_{,t}}{|\bm{\beta}|L^2} \bigg] \quad L^2 d\hat{y} d\hat{z} \nonumber\\ \nonumber\\
\int_{-L/2}^{L/2} \Phi_{, t}|_{x=L/2} \ dy dz &= G ML_{,t} \int_{-1/2}^{1/2} \bigg[ \lim _{\mathcal{N} \to \infty} \sum_{\bm{\beta} = - \mathcal{N}}^{\mathcal{N}}\frac{ ( \beta_{1}(1/2)+\beta_{2}\hat{y}+\beta_{3}\hat{z} - |\bm{\beta}|^2)}{[(1/2 -\beta_{1})^2 + (\hat{y} -\beta_{2})^2 + (\hat{z}-\beta_{3})^2]^{3/2}} +   \lim _{\mathcal{N} \to \infty} \sum_{\bm{\beta^{*}} = - \mathcal{N}}^{\mathcal{N}}\frac{1}{|\bm{\beta}|} \bigg] \ d\hat{y} d\hat{z}  \nonumber\\
 &\equiv G M EL_{,t}  \ ,  \label{117} %= 2G ME X^{(0)}_{,t}
\end{align}
where the last line gives the definition of $E$. Finally, the second time derivative of $\Phi$, integrated over the volume, is
\begin{align}
&\int_{-L/2}^{L/2} \Phi_{, t t} \ dx dy dz \nonumber \\=&  \int_{-L/2}^{L/2} \bigg[ \lim _{\mathcal{N} \to \infty} \sum_{\bm{\beta} = - \mathcal{N}}^{\mathcal{N}}\frac{3G M( \beta_{1}x+\beta_{2}y+\beta_{3}z - |\bm{\beta}|^2 L)^2 L_{,t}^2  }{|\mathbf{x} - \bm{\beta}L|^5}   +  \lim _{\mathcal{N} \to \infty} \sum_{\bm{\beta} = - \mathcal{N}}^{\mathcal{N}}\frac{G M( \beta_{1}x+\beta_{2}y+\beta_{3}z - |\bm{\beta}|^2 L) L_{,tt} }{|\mathbf{x} - \bm{\beta}L|^3} \nonumber \\ &\, \hspace{4.5cm}- \lim _{\mathcal{N} \to \infty} \sum_{\bm{\beta} = - \mathcal{N}}^{\mathcal{N}}\frac{|\bm{\beta}|^2 G M{L_{,t}}^2}{|\mathbf{x} - \bm{\beta}L|^3}  +  \lim _{\mathcal{N} \to \infty} \sum_{\bm{\beta^{*}} = - \mathcal{N}}^{\mathcal{N}}\frac{G ML_{,tt}}{|\bm{\beta}|L^2} -   \lim _{\mathcal{N} \to \infty} \sum_{\bm{\beta^{*}} = - \mathcal{N}}^{\mathcal{N}}\frac{2G M{L_{,t}}^2}{|\bm{\beta}|L^3}  \bigg] \ dx dy dz  \nonumber\\  
 =& G ML_{,t}^2  \int_{-1/2}^{1/2} \bigg[ \lim _{\mathcal{N} \to \infty} \sum_{\bm{\beta} = - \mathcal{N}}^{\mathcal{N}}\frac{3 ( \beta_{1}\hat{x}+\beta_{2}\hat{y}+\beta_{3}\hat{z} - |\bm{\beta}|^2)^2 }{|\mathbf{ \hat{x}} - \bm{\beta}|^5} - \lim _{\mathcal{N} \to \infty} \sum_{\bm{\beta} = - \mathcal{N}}^{\mathcal{N}}\frac{|\bm{\beta}|^2}{|\mathbf{\hat{x}} - \bm{\beta}|^3} -   \lim _{\mathcal{N} \to \infty} \sum_{\bm{\beta^{*}} = - \mathcal{N}}^{\mathcal{N}}\frac{2}{|\bm{\beta}|}\bigg] \ d\hat{x} d\hat{y} d\hat{z}  \nonumber\\  
&+ G MLL_{,tt} \int_{-1/2}^{1/2} \bigg[ \lim _{\mathcal{N} \to \infty} \sum_{\bm{\beta} = - \mathcal{N}}^{\mathcal{N}}\frac{( \beta_{1}\hat{x}+\beta_{2}\hat{y}+\beta_{3}\hat{z} - |\bm{\beta}|^2) }{|\mathbf{ \hat{x}} - \bm{\beta}|^3} +  \lim _{\mathcal{N} \to \infty} \sum_{\bm{\beta^{*}} = - \mathcal{N}}^{\mathcal{N}}\frac{1}{|\bm{\beta}|}\bigg] \ d\hat{x} d\hat{y} d\hat{z} \ \nonumber\\ 
\equiv& G ML_{,t}^2 F +  G MLL_{,tt} P \ ,  \label{118}
\end{align}
where $F$ and $P$ are defined by the last line.
\end{widetext}

For a single point source at the centre of a cell we take $v^{\alpha} = p = \Pi =\avg{\rho \Phi} =0$. Making use of Eqs. \eqref{116}, \eqref{117} and \eqref{118}, as well as the lower-order solutions, we then find that Eq. \eqref{94} reduces to
\begin{align} \label{120}
 X_{,tt} &= -\frac{GM}{6\zeta^2} \bigg[ \pi + 5 \pi C - {9} E C - {3} F C  \bigg] \nonumber \\ 
 & \quad + \frac{\pi G^2 M^2}{6(X^{(0)})^3} \bigg[ 2 D  + \frac{P}{2} -  F - 3 E +  \frac{14\pi}{9}\bigg]   \nonumber \\
&\quad  - \bigg(\frac{7\pi^2 G^2 M^2}{36 (X^{(0)})^5}  -\frac{\pi G M C}{2(X^{(0)})^4} \bigg)(y^2 +z^2)+ O(\epsilon^6)  \ . 
\end{align}

%D= 1.44, E = 0.643, F = -1.63, P = 0.304
To solve this equation, we can evaluate it at the centre of the cell face, where $y=0$ and $z=0$. In this case, \\ $\zeta = X$ and it is convenient to recombine the terms involving $X^{(0)}$ and $X^{(2)}$, to find
\begin{align} 
 X_{,tt} &= -\frac{G M}{6 X^2} \bigg[ \pi + 5 \pi C - {9} E C - {3} F C  \bigg]   \nonumber \\ 
 & \quad + \frac{\pi G^2 M^2}{6X^3} \bigg[2 D  + \frac{P}{2} -  F - 3 E +  \frac{14\pi}{9}\bigg]  + O(\epsilon^6) \ . \nonumber \\ 
% X_{,tt} &\approx - \frac{1}{24X^2} \bigg[4\pi G M + 22\pi G M C + 4\pi G \avg{\rho \Phi} + 4\pi G \avg{\rho \Pi} +36 G M E C + 12 G M F C  \bigg]  \nonumber \\ 
% & \quad + \frac{1}{24X^3} \bigg[31.27 \pi G^2 M^2 \bigg] + O(\epsilon^6) \ . \nonumber \\ 
  &\equiv - \frac{N}{X^2} + \frac{J}{X^3}  + O(\epsilon^6) \ , \label{121}
\end{align}
where the last line defines $N$ and $J$. 

\begin{figure}[t!]
\includegraphics[width=0.52 \textwidth]{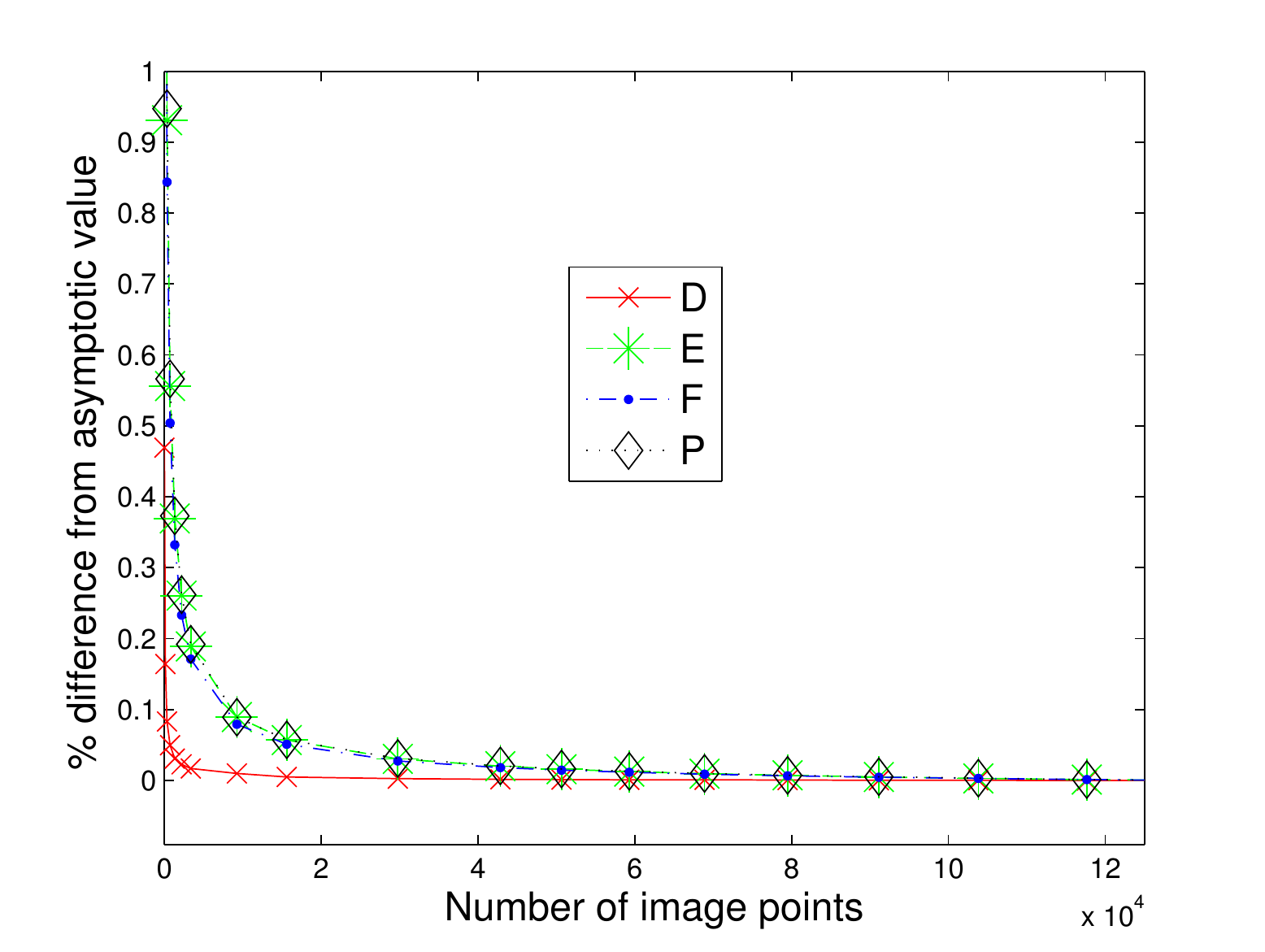}
\caption{\label{fig4} The percentage difference from the asymptotic value of $D, E, F$ and $P$, for various numbers of image points in the partial sum.}
\end{figure}

As can be seen from Table \ref{tab2}, and the plots in Figure \ref{fig4}, the numerical constants $D, E, F$ and $P$ are all of order unity, and converge rapidly as the number of image masses becomes large. The two terms on the right-hand side of Eq. \eqref{121} look like dust and radiation, respectively. However, $J = 1.27\pi G^2 M^2$ is positive, so the radiation-like term appears to have negative energy density, and hence contributes positively to the acceleration of the boundary. 
 %X_{,t}X_{,tt}&= \frac{N X_{,t}}{X^2} + \frac{J X_{,t}}{X^3}  \ , \nonumber \\ \nonumber \\
%\frac{1}{2} (X_{,t}^2)_{,t}&= \frac{N X_{,t}}{X^2} + \frac{J X_{,t}}{X^3}   \ , \nonumber \\ \nonumber \\
% X_{,t}^2 &= 2\int \frac{N X_{,t}}{X^2} + \frac{J X_{,t}}{X^3} \ dt  \ , \nonumber \\ \nonumber \\

Integrating Eq. \eqref{121} gives the Friedmann-like equation
\begin{align}
 X_{,t}^2 &=  \frac{2N}{X} -\frac{J}{X^2} - C  + O(\epsilon^6)\, ,  \label{122}
 \end{align}
where $C$ is a constant. The solutions to Eq. \eqref{122} depend on the value of $C$. If $C=0$ then
\begin{align} \label{123}
X= \frac{N}{2} \eta^2 + \frac{J}{2N} \ ,\nonumber \\  \nonumber \\
t -t_{0} = \frac{J}{2N} \eta + \frac{N}{6}\eta^3 \ .
\end{align}
If $0<C< \frac{N^2}{J}$ then
\begin{align} \label{124}
X= \frac{N}{C} \pm \frac{1}{C}\sqrt{N^2 - JC} \sin\big[\sqrt{C}\eta \big] \ ,\nonumber \\  \nonumber \\
t -t_{0} = \frac{N}{C}\eta \mp \frac{1}{C^{3/2}}\sqrt{N^2 - JC} \cos\big[\sqrt{C} \eta \big] \ .
\end{align}
In both of these equations $\eta = \int dt/X$ is analogous to the conformal time parameter used in Friedmann cosmology.

\begin{figure}[t!]
\includegraphics[width=0.52 \textwidth]{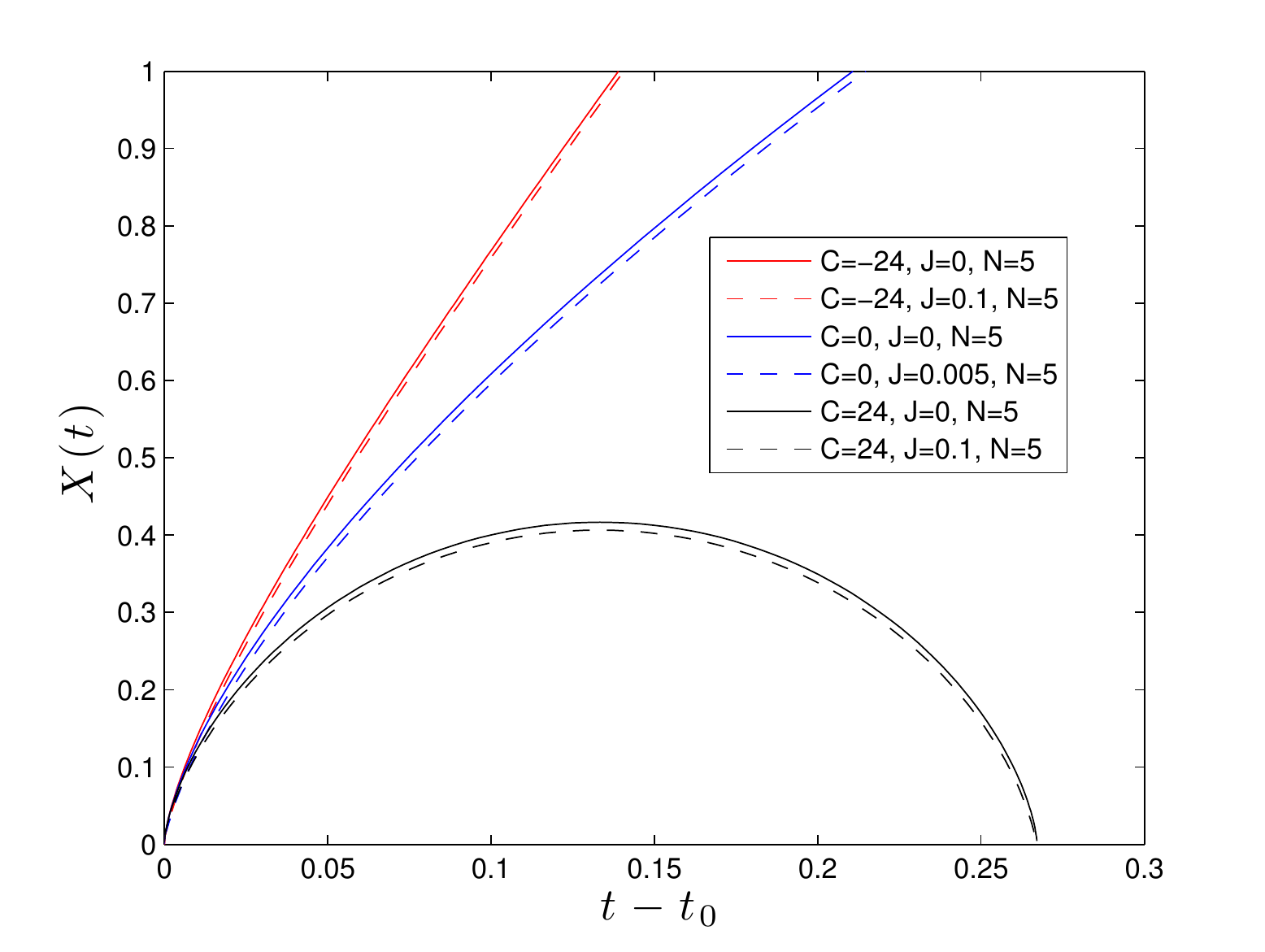}
\caption{\label{fig5} Illustrations of the solutions of $X(t)$ for different values of $C$ and $J$. The constant J will likely take much smaller values, for realistic configurations. Its value is exaggerated here to illustrate the effect of the post-Newtonian terms.}
\end{figure}

For $C<0$, on the other hand, we obtain
\begin{align} 
&\pm (t -t_{0}) \nonumber \\=& \frac{N}{(-C)^{3/2}} \ln \bigg| -N + CX + \sqrt{-C}\sqrt{-CX^2 + 2NX - J}\bigg|\nonumber \\&+\frac{\sqrt{-CX^2 + 2NX - J}}{(-C)}   \, ,\label{125}
\end{align} 
where we have written the inverted function $t(X)$, for convenience. The arguments under the square root must be positive for there to be real solutions, which is always true if $J>0$ and $C<0$. For $C\geqslant \frac{N^2}{J}$ there are no real solutions. In Figure \ref{fig5} we present the functional form of $X(t)$ for three example values of $C$ and $J$.

\begin{table}[b!]
\begin{tabular}{ | c | l |  }
    \hline 
    \textbf{\, Constant \,} & \textbf{\, Asymptotic Value \,} \\ \hline 
    $D$ & $\qquad \phantom{-}1.44 \ldots$   \\ \hline %radius of insphere X - a\ \sqrt{24}
    $E$ & $\qquad \phantom{-}0.643 \dots$ \\  \hline
    $F$ & $\qquad -1.63 \dots$  \\ \hline
    $P$ & $\qquad \phantom{-}0.304 \dots$ \\ \hline
\end{tabular}
  \caption{\label{tab2} The asymptotic values of $D$, $E$, $F$ and $P$, which are approached as the number of image masses diverges to infinity.} 
\end{table}

If we were to extrapolate these solutions beyond their reasonable regime of validity, to very early times, then we would observe a bounce at the following minimum values of $X$:
\begin{equation} \label{126}
X_{min}= \begin{cases} \frac{N}{C} - \frac{\sqrt{N^2-JC}}{C}, & \mbox{if } C\neq0 \\ \frac{J}{2N}, & \mbox{if } C=0 \ . \end{cases}  
\end{equation}
This concludes our discussion of explicit solutions for $X(t)$, in the presence of point-like particles.

\section{Relationship with FLRW models, and observables} \label{sec7}

So far, we have only calculated coordinate distances, in terms of coordinate time. In this section we will transform these quantities into proper distances and proper time. We will then perform a coordinate transformation that allows the space-time within each cell to be written as perturbations on a homogeneous and isotropic Robertson-Walker background. Finally, we will work out the expansion rates of our cell edges in this new description. This will allow us the clearest possible comparison with standard Friedmann-Lema\^{i}tre-Robertson-Walker (FLRW) cosmological models. As an example, we will sometimes  use the model described in Section \ref{sec6b}. Similar analyses can be performed for other configurations, by adjusting the calculations that follow.

\subsection{The Proper Length of Cell Edges} \label{sec7a}

For our example cubic cell, we choose to consider an edge defined by the intersection of cell faces at $x=X(t,y,z)$ and $y=Y(t,x,z)$. The proper length of such a curve, in a hypersurface of constant $t$, is given by
\begin{align}
\mathcal{L} = \int_{\rm edge} \sqrt{(1+2 \Phi) (dx^2+dy^2+dz^2)} +O(\epsilon^4) \, .
\end{align}
Expanding this expression, and taking the locations of the relevant corners of the cell to be at $z= \pm Z_c$, then gives
\begin{align} \label{127}
 \mathcal{L} %&= \int_{-Z_c}^{Z_c} \sqrt{g_{zz}} \ dz 
 =  \int_{-Z_c}^{Z_c} (1+\Phi) \ dz +O(\epsilon^4) \, .
\end{align}
Using Eqs. \eqref{60} and \eqref{84}, it can be seen that
\begin{equation} \label{127b}
Z_c = \frac{L}{2} - \frac{\pi G M}{6}  +O( \epsilon^4) \, ,
\end{equation}
where $L$ is being used here to denote the coordinate distance between the centres of two cell faces, on opposite sides of our cubic cell.

Using the solution derived in Section \ref{sec6b}, we can numerically evaluated the integral in Eq. \eqref{127}, between the limits specified in Eq. \eqref{127b}, to find
\begin{align}
 \mathcal{L} 
%&\simeq \int_{-L/2}^{L/2} (1+\Phi) \ dz +O(\epsilon^4)  \nonumber \\ \nonumber %\\
%  &= L +  \int_{-L/2}^{L/2} \bigg[ \sum_{\bm{\beta}\in \mathbb{Z}^3}\frac{GM}{\sqrt{(L/2 %-\beta_{1}L)^2 + (L/2 -\beta_{2}L)^2 + (z-\beta_{3}L)^2 }} -  \sum_{\bm{\beta}\in %\mathbb{Z}_{*}^3}\frac{M}{|\bm{\beta}|L}\bigg] \ dz +O(\epsilon^4)  
%\nonumber \\
%  &\simeq GM \int_{-1/2}^{1/2} \sum_{\bm{\beta}\in \mathbb{Z}^3}\frac{1}{\sqrt{(1/2-\beta_{1})^2 + (1/2 -\beta_{2})^2 + ({z} -\beta_{3})^2 }} dz -\int_{-1/2}^{1/2}  \sum_{\bm{\beta}\in \mathbb{Z}_{*}^3}\frac{1}{|\bm{\beta}|}\ \ d{z} \nonumber\\
%  &+  \nonumber \\
%  &               \nonumber \\
%  &\simeq L + 0.922GM  
\simeq L - 0.125 GM \, . \label{128}
\end{align}
Using this result, the Friedmann-like equation \eqref{122} can then be written as
\begin{align} \label{129}
 %\mathcal{L} _{,t}^2 &\simeq  \frac{16N}{ \mathcal{L}}+ \frac{16 N \times 0.922G M}{\mathcal{L}^2} -\frac{16J}{ \mathcal{L}^2} - 4C  + O(\epsilon^6)\nonumber \\ \nonumber \\
% \mathcal{L} _{,t}^2 &\simeq  \frac{16N}{ \mathcal{L}}+ \frac{7.38\pi G^2 M^2}{3\mathcal{L}^2} -\frac{16J}{ \mathcal{L}^2} - C  + O(\epsilon^6)\ .
\mathcal{L} _{,t}^2 &\simeq  \frac{16N}{ \mathcal{L}} -\frac{64.9 G^2 M^2}{ \mathcal{L}^2} - 4C   \, ,
\end{align}
The functional form of this equation is obviously unaltered, with only a small additional contribution to the radiation-like term.

Let us now write this equation in terms of proper time, $\tau$, of an observer following the boundary at the corner of the cell. In this case coordinate time can be related to proper time using the normalisation of the 4-velocity tangent to the boundary, $U^{a}U_{a} = -1$, so that
\begin{align} \label{130}
U^{t} &= \frac{d t}{d \tau} = 1 + \Phi + \frac{3}{2} {(X_{,t})}^2 +O(\epsilon^4) \, ,
\end{align}
where all terms are to be evaluated at the corner of the cell, and where the observer has been taken to be moving with velocity equal to $\pm X_{,t}$ in each of the $x$, $y$ and $z$-directions. This equation can be re-written using the solution in Section \ref{sec6b}, to find 
\begin{align}
%& \simeq  1+ \frac{1.11GM}{X} +  \frac{\pi G M}{6X} - C +O(\epsilon^4) \nonumber \\ \nonumber \\
%& \simeq  1+ \frac{2.22GM}{ \mathcal{L}} +  \frac{\pi G M}{3 \mathcal{L}} - C +O(\epsilon^4) \ ,
%U^t \simeq  1+ \frac{5.36 GM}{ \mathcal{L}} - \frac{3C}{2}  \, , \nonumber \\
U^t \simeq  1+ \frac{3.61GM}{ \mathcal{L}} - \frac{3C}{2}  \, ,
\end{align}
where we have used Eq. \eqref{128} to replace coordinate distances with proper length. In terms of this proper time, Eq. \eqref{129} is given by
\begin{align} 
% \mathcal{L} _{,\tau}^2 &\simeq  \frac{16N}{ \mathcal{L}}- \frac{16\pi GMC}{3 \mathcal{L}}+\frac{35.52\pi G^2 M^2}{3\mathcal{L}^2}  +\frac{16\pi^2 G^2 M^2}{9\mathcal{L}^2}+\frac{7.38\pi G^2 M^2}{3\mathcal{L}^2} -\frac{16J}{ \mathcal{L}^2} - 4C  + O(\epsilon^6)  \nonumber \\ \nonumber \\
%  \mathcal{L} _{,\tau}^2 &\simeq  \frac{16N}{ \mathcal{L}} - \frac{16\pi GMC}{3 \mathcal{L}} +\frac{19.89\pi G^2 M^2}{\mathcal{L}^2}  -\frac{16J}{ \mathcal{L}^2} - 4C  + O(\epsilon^6)  \nonumber \\ \nonumber \\
%   \mathcal{L} _{,\tau}^2 &\simeq  \frac{16N}{ \mathcal{L}} - \frac{16\pi GMC}{3 \mathcal{L}} +\frac{62.5 G^2 M^2}{\mathcal{L}^2}  -\frac{16J}{ \mathcal{L}^2} - 4C + 8C^2 - \frac{6.54 \times 4 GMC}{\mathcal{L}}  + O(\epsilon^6)  \nonumber \\ \nonumber \\
 %&\simeq  \frac{16N}{ \mathcal{L}} - \frac{16\pi GMC}{3 \mathcal{L}} -\frac{0.91\pi G^2 M^2}{\mathcal{L}^2} - 4C + 8C^2  + O(\epsilon^6)   \ ,  \label{131}
%\hspace{-0.25cm} \left( \frac{d \mathcal{L}}{d \tau} \right)^2 &\simeq  \frac{(16N-68.0 GMC)}{ \mathcal{L}} +\frac{23.4 G^2 M^2}{\mathcal{L}^2} - 4C (1-3 C)   \, .   \nonumber \\
\hspace{-0.25cm} \left( \frac{d \mathcal{L}}{d \tau} \right)^2 &\simeq  \frac{(16N-54.0 GMC)}{ \mathcal{L}} -\frac{4.41 G^2 M^2}{\mathcal{L}^2} - 4C (1-3 C)   \, .  \label{131}
\end{align}
The functional form of this equation is again the same as Eq. \eqref{122}, but now with corrections to both the dust-like and spatial curvature-like terms, as well as the radiation-like term. 

\subsection{Transformation to a Time-dependent Background} \label{sec7b}

Up to this point we have treated the geometry within each cell as a perturbation to Minkowski space. This is convenient, as it allows the apparatus constructed for the study of post-Newtonian gravity in isolated systems to be applied with only minimal modifications. When considering the case of a cosmological model, however, it is also of use to be able to understand the form of the gravitational fields in the background of a Robertson-Walker geometry. In this section we will show that the perturbed Minkowski space description and the perturbed Robertson-Walker description are isometric to each other, as long as we restrict our consideration to a single cell. We will present coordinate transformations that make this isometry explicit. This discussion follows, and extends, that presented in \cite{Tim1}.

We begin by writing the unperturbed line-element for a Robertson-Walker geometry as
\begin{equation}
ds^{2} = -d\hat{t}^2 + a(\hat{t})^2 \frac{ \left( d \hat{x}^2 + d \hat{y}^2 + d \hat{z}^2 \right)}{[1+\frac{k}{4} (\hat{x}^2 + \hat{y}^2 + \hat{z}^2)]^2} \, ,
\end{equation}
where $a(\hat{t})$ and $k$ are the scale factor and curvature of hypersurfaces of constant $\hat{t}$, respectively. From the Friedmann equations we know that $k \sim (a_{,\hat{t}})^2 \sim a_{,\hat{t} \hat{t}}$. As $a \sim 1$, this means that we must immediately require that $k \sim \epsilon^2$, which means that $k$ can be treated as a (homogeneous) perturbation to a spatially-flat background.

With this information, we can write the line-element of a perturbed Robertson-Walker geometry as
\begin{align}  
ds^{2} 
\simeq& -(1 - \hat{h}^{(2)}_{\hat{t} \hat{t}} -  \hat{h}^{(4)}_{\hat{t} \hat{t}})d\hat{t}^2 
+  2a(\hat{t}) \hat{h}^{(3)}_{ \hat{t}\hat{\mu}} dx^{\hat{\mu}} d\hat{t}
\nonumber \\ &+  a(\hat{t})^2 \bigg(\delta_{\hat{\mu}\hat{\nu}}+\hat{h}^{(2)}_{\hat{\mu} \hat{\nu}}- \frac{k}{2} \delta_{\hat{\mu}\hat{\nu}} (\hat{x}^2 + \hat{y}^2 + \hat{z}^2)\bigg) dx^{\hat{\mu}} dx^{\hat{\nu}}  \, ,  \label{132}
\end{align}
where the hatted coordinates $x^{\hat{\mu}} =(\hat{x},\hat{y},\hat{z})$, and where $\hat{h}_{\hat{a}\hat{b}}$ are perturbations. In this expression we have written the $\hat{t}\hat{t}$-component of the metric to $O(\epsilon^4)$, the $\hat{\mu} \hat{\nu}$-component to $O(\epsilon^2)$, and the $\hat{t}\hat{\mu}$-component to $O(\epsilon^3)$.

The lowest-order part of the Einstein's field equations give, for the perturbed metric in Eq. \eqref{132},
\begin{equation}
\hat{h}^{(2)}_{\hat{\mu} \hat{\nu}} = \hat{h}^{(2)}_{\hat{t} \hat{t}} \delta_{\hat{\mu}\hat{\nu}} \, ,
\end{equation}
and
\begin{align}
8\pi G \hat{\rho}a^3 + a\hat{\nabla}^2 \hat{h}^{(2)}_{\hat{t} \hat{t}} = -6 a_{,\hat{t}\hat{t}}a^2 = 3 (a_{,\hat{t}})^2 a +  3 k a\, .  \label{133}
\end{align}
An integrability condition of this latter equation is that
\begin{equation}
8\pi G \hat{\rho}a^3 +a\hat{\nabla}^2 \hat{h}^{(2)}_{\hat{t} \hat{t}} = {\rm constant} \equiv 2 C_4 \, ,
\end{equation}
where by constant we mean not a function of $\hat{t}, \hat{x}$, $\hat{y}$ or $\hat{z}$. This result shows that the $\nabla^2 \hat{h}^{(2)}_{\hat{t} \hat{t}}$ term in Eq. \eqref{133} behaves like dust, in the way that it sources the evolution of the scale factor.

Let us now consider the following coordinate transformations:
\begin{align} 
\hat{t} &= t - \frac{a_{,t}}{2a} (x^2 + y^2 + z^2) + T(t,x^{\mu}) +  O(\epsilon^5) \nonumber \\ \nonumber \\
\hat{x} &= \frac{x}{a}\bigg[1 + \frac{(a_{,t})^2}{4a^2} (x^2 + y^2 + z^2)\bigg] +O(\epsilon^4)   \nonumber \\ \nonumber \\
\hat{y} &= \frac{y}{a}\bigg[1 + \frac{(a_{,t})^2}{4a^2} (x^2 + y^2 + z^2)\bigg] +O(\epsilon^4)   \nonumber \\ \nonumber \\
\hat{z} &= \frac{z}{a}\bigg[1 + \frac{(a_{,t})^2}{4a^2} (x^2 + y^2 + z^2)\bigg] +O(\epsilon^4) \, ,  \label{135}
\end{align}
where $T(t, x^{\mu})$ is an unspecified quantity of $O(\epsilon^3)$. The scale factor on the right-hand side of these equations is written as a function of the time coordinate $t$, and is related to $a(\hat{t})$ by
\begin{equation}
a(\hat{t}) = a(t)\bigg[1 - \frac{(a_{,t})^2}{2a^2} (x^2 + y^2 + z^2)\bigg] +O(\epsilon^4) \, .
\end{equation}
In the un-hatted coordinates, the perturbed metric in Eq. \eqref{132} transforms into the one given in Eq. \eqref{4}, where the perturbations around the Robertson-Walker geometry are given in terms of the perturbations about Minkowski space in the following way:
\begin{align}
\hat{h}^{(2)}_{\hat{t} \hat{t}} &= {h}^{(2)}_{{t} {t}}  -  \frac{a_{,tt}}{a} (x^2 + y^2 + z^2) +O(\epsilon^4)  \nonumber \\
\hat{h}^{(3)}_{ t\mu} &= h^{(3)}_{t\mu} + T_{,\mu} + 2\frac{a_{,t}}{a}x^{\mu}{h}^{(2)}_{{t} {t}} \nonumber \\&\quad +\bigg(\frac{2C_{4}}{3a} - \frac{k}{2}\bigg) \frac{a_{,t}}{a^3} x^{\mu}  (x^2 + y^2 + z^2) +O(\epsilon^5)\nonumber \\
%\hat{\Phi}^{(4)}&= \Phi^{(4)} + T_{,t} + \frac{a_{,t}}{a} \hat{h}_{t\mu} x^{\mu} + \frac{5C_{4}}{3 a^3}\Phi(x^2 +y^2 +z^2) - \frac{{C_{4}}^2}{24 a^6} (x^2 +y^2 +z^2)^2  
 %\hat{h}^{(4)}_{\hat{t} \hat{t}}&=h_{tt} + 2T_{,t} + 2\frac{a_{,t}}{a} \hat{h}_{t\mu} x^{\mu} \nonumber \\ &\quad - \bigg(\frac{5C_{4}}{3 a^3} + \frac{2k}{a^2}\bigg){h}^{(2)}_{{t} {t}} (x^2 +y^2 +z^2) \nonumber \\ &\quad - \frac{3}{4}\bigg( \frac{{C_{4}}^2}{a^6} + \frac{k^2}{a^4} - \frac{2k C_{4}}{ a^5} \bigg) (x^2 +y^2 +z^2)^2  +O(\epsilon^6)  \nonumber \\
% &=h_{tt} + 2T_{,t} + 2\frac{a_{,t}}{a} \hat{h}_{t\mu} x^{\mu} - \bigg(\frac{10C_{4}}{3 a^3} + \frac{4k}{a^2}\bigg)\Phi (x^2 +y^2 +z^2) - \frac{3}{4}\bigg( \frac{{C_{4}}^2}{a^6} + \frac{k^2}{a^4} - \frac{2k C_{4}}{ a^5} \bigg) (x^2 +y^2 +z^2)^2  +O(\epsilon^6) \nonumber \\ \nonumber \\
% & \quad + 2\frac{a_{,t}}{a} (h_{t\mu} + T_{,\mu})x^{\mu} + 8\frac{a_{,t}^2}{a^2}(x^{\mu}x_{\mu})\Phi +\bigg(\frac{2C_{4}}{3a} - \frac{k}{2}\bigg) \frac{2{a_{,t}}^2}{a^4} (x^{\mu}x_{\mu})^2) \nonumber \\ \nonumber \\
% & \quad = 2\frac{a_{,t}}{a} (h_{t\mu} + T_{,\mu})x^{\mu} + 8\bigg(\frac{2C_{4}}{3a^3} - \frac{k}{a^2}\bigg)(x^{\mu}x_{\mu})\Phi +\bigg(\frac{4C_{4}}{3a^3} - \frac{k}{a^2}\bigg)\bigg(\frac{2C_{4}}{3a^3} - \frac{k}{a^2}\bigg) (x^{\mu}x_{\mu})^2) \nonumber \\ \nonumber \\
   \hat{h}^{(4)}_{\hat{t} \hat{t}}&=h^{(4)}_{tt} + 2T_{,t} + 2\frac{a_{,t}}{a} (h^{(3)}_{t\mu} + T_{,\mu})x^{\mu} \nonumber \\ &\quad + \bigg(\frac{C_{4}}{a^3} - \frac{6k}{a^2}\bigg){h}^{(2)}_{{t} {t}} (x^2 +y^2 +z^2) \nonumber \\ &\quad + \bigg( \frac{5{C_{4}}^2}{36a^6} - \frac{k C_{4}}{ 2a^5} + \frac{k^2}{4a^4} \bigg) (x^2 +y^2 +z^2)^2  +O(\epsilon^6) \, .\label{136}
\end{align}
These equations cannot be used to transform a global perturbed Robertson-Walker geometry to a global perturbed Minkowski geometry, as the velocities that result in the latter would be greater than the speed of light on scales of $H_0^{-1}$. They are, however, perfectly sufficient to transform the geometry within any one of our cells. As every cell is identical, this provides us with a way to transform the entire geometry.

We can use these transformations to relate the proper length calculated in the Minkowski space background, $\mathcal{L}$, to the proper length in a flat Robertson-Walker background, $\hat{\mathcal{L}}$. The relevant corners in this background are given by $\hat{z}=\pm\hat{Z_c}$, which can be related to the positions of the cell corners in the Minkowski space background using Eq. \eqref{135}, so that
\begin{align} \label{hatcorner}
\hat{Z_c} &= \frac{Z_c}{a} + \frac{3(a_{,t})^2 X^3}{4a^3} +O(\epsilon^4)
\end{align}
For a flat Robertson-Walker background we do not need any $O(\epsilon^2)$ corrections to $\hat{Z_c}$, as we did for $Z_c$ in Eq. \eqref{127b}, because the boundaries are flat to this order \cite{Tim1}.

Using Eqs. \eqref{135} and \eqref{136}, $\hat{\mathcal{L}}$ is given by
\begin{align}
 \hat{\mathcal{L}} &\simeq \int_{-\hat{Z_c}}^{\hat{Z_c}} \hat{a} \bigg(1+\hat{\Phi} 
% - \frac{k}{4} (\hat{x}^2 + \hat{y}^2 + \hat{z}^2)
 \bigg) \ d\hat{z} \nonumber \\
&\simeq  \int_{-\hat{Z_c}}^{\hat{Z_c}} a \bigg(1 - \frac{(a_{,t})^2}{4a^2} (x^2 + y^2 + z^2)+\Phi \bigg) \ d\hat{z}  \, . \label{138}
\end{align}
Using $z =a\hat{z}$ at lowest order, and again choosing the edge of our cell at $x=X$ and $y=Y$, we can integrate part of this equation to find 
\begin{align}
     \hat{\mathcal{L}} &\simeq \int_{-\hat{Z_c}}^{\hat{Z_c}} \bigg(a - \frac{a(a_{,t})^2}{4} \bigg(\hat{z}^2 + \frac{L^2}{2a^2}\bigg) \bigg)  \ d\hat{z} + \int_{-Z_c}^{Z_c} a\Phi  \ \frac{dz}{a} +O(\epsilon^4) \nonumber \\ \nonumber \\
 &\simeq 2\hat{Z_c}a - \frac{a(a_{,t})^2}{4} \bigg(\frac{2\hat{Z_c}^3}{3} + \frac{L^2\hat{Z_c}}{a^2}\bigg) + \int_{-Z_c}^{Z_c} \Phi  \ dz  \ . \label{139}
\end{align}
The last term in this equation can be evaluated numerically, in the same way we evaluated the last term in Eq. \eqref{128}. We will do this below, for the explicit solution found in Section \ref{sec6b}, consisting of a single point-like mass at the centre of every cell.

Using Eqs. \eqref{hatcorner} and \eqref{139} we then obtain
\begin{align}
  \hat{\mathcal{L}} %&\simeq 2X + \frac{2a(a_{,t})^2}{4} X^3 -\frac{2 a(a_{,t})^2}{12} X^3 - \frac{L^2 X a(a_{,t})^2}{4a^3} + 0.922 GM + O(\epsilon^4) \nonumber \\ \nonumber \\
%&\simeq L - \frac{2 X^3 (a_{,t})^2}{3 a^2} + 0.922 GM  \nonumber \\
 %&\simeq \mathcal{L} - \frac{4 X^3 C_{4}}{9 a^3} + 1.047 GM \nonumber \\ %+ \frac{2 X^3 k }{3 a^2}  
 &\simeq L - \frac{\pi G M}{3} - \frac{5 X^3 (a_{,t})^2}{12 a^2} + 0.922 GM  \nonumber \\
 &\simeq \mathcal{L} - \frac{5 X^3 C_{4}}{18 a^3} \ , \label{140}
\end{align}
where we have used Eqs. \eqref{128} and \eqref{133} in the last line. For a flat FLRW background we can use Eq. \eqref{133}, together with the solution for the scale factor, to find
\begin{align}
 a(t)= \bigg(\frac{3}{2}\bigg) ^{2/3} \left(\sqrt{\frac{2C_{4}}{3}} t - t_{0} \right)^{2/3} \, , \label{134}
\end{align}
where $C_{4}$ has been taken to be positive. Now, using Eqs. \eqref{65} and \eqref{134}, for the solutions of $X$ and $a$, respectively, we can infer that
\begin{align}
     C_{4} = \frac{\pi G M a^3}{2X^3} \, . \label{141}
\end{align}
The proper length of a cell edge in the Robertson-Walker background is therefore given by
\begin{align}
        \hat{\mathcal{L}} %&= \mathcal{L} - \frac{2\pi GM}{9} + O(\epsilon^4) \nonumber \\ 
       %&\simeq \mathcal{L} +0.349GM  \nonumber \\ \,  
        %&\simeq \mathcal{L} +0.611GM  - \frac{\pi G M}{3} \nonumber \\
         &\simeq  \mathcal{L} - 0.436GM \ .\label{142}
\end{align}
The Friedmann-like equation can then be written in terms of this quantity as
\begin{align}
\hspace{-0.35cm} \left(\frac{d\hat{\mathcal{L}}}{d\tau} \right)^2 %&\simeq \frac{16N}{  \hat{\mathcal{L}} + 0.698 G M} - \frac{42.9 GMC}{\hat{\mathcal{L}}} -\frac{2.86 G^2 M^2}{ \hat{\mathcal{L}}^2}- 4C + 8C^2  + O(\epsilon^6)   \nonumber \\ \nonumber \\
  %  &=  \frac{16N}{  \hat{\mathcal{L}} } \bigg(1 + \frac{0.698 G M}{ \hat{\mathcal{L}} } \bigg)^{-1}- \frac{42.9 GMC}{\hat{\mathcal{L}}} -\frac{2.86 G^2 M^2}{ \hat{\mathcal{L}}^2}- 4C + 8C^2   + O(\epsilon^6)   \nonumber \\ \nonumber \\
    % &=  \frac{16N}{  \hat{\mathcal{L}} } - \frac{5.85G^2 M^2}{ \hat{\mathcal{L}}^2 } - \frac{42.9 GMC}{\hat{\mathcal{L}}} -\frac{2.86 G^2 M^2}{ \hat{\mathcal{L}}^2} - 4C + 8C^2  + O(\epsilon^6)   \nonumber \\ \nonumber \\
      %&\simeq  \frac{(16N-68.0 GMC)}{  \hat{\mathcal{L}} } +\frac{26.32 G^2 M^2}{ \hat{\mathcal{L}}^2} - 4C (1-3C)   \nonumber \\ 
       &\simeq  \frac{(16N-54.0 GMC)}{  \hat{\mathcal{L}} } - \frac{8.06 G^2 M^2}{ \hat{\mathcal{L}}^2} - 4C (1-3C)    \, . \label{143}
\end{align}
This equation has the same form as Eq. \eqref{131}, but with a different numerical coefficient for the radiation-like term. Again, we remind the reader that this term, although it takes the form of a radiation fluid in the evolution equation for the scale of the space, does not correspond to any actual matter field. It is purely a result of the non-linearity of Einstein's equations.

\subsection{Observables}

The majority of this paper has been aimed at constructing a geometry to model the space-time of a universe that can accommodate large non-linear density contrasts on small scales, while maintaining a periodic structure that means it is invariant under a certain set of discrete spatial translations on large scales. Space-time geometry itself, however, is not a direct observable. We will therefore pause, and consider in this section the consequences of this new type of model for cosmological observables.

We will briefly consider two types of observations: (i) those based on the motion of time-like particles (i.e. galaxies, dark matter, etc.), and (ii) those based on observations made along null geodesics. Of course, almost all observations are in fact made by observing light, but one can in principle split ones considerations into the effect of the expansion of space on the light directly, and the effect of the expansion on the astrophysical objects that emit or interact with the light.

Let us start with the motion of time-like particles. The model constructed above, and in particular the Green's function formalism adopted in Section \ref{sec6a}, allows $n$-body simulations of structure formation to be modelled to post-Newtonian accuracy in a fully self-consistent way. This problem has recently been studied using a different approach in \cite{nbody2}, where actual simulations were performed, and observables extracted. The effects of post-Newtonian gravitational fields are found to be small in all aspects of these simulations, but are nonetheless important for understanding the consequences of relativistic gravity, and in order to accurately interpret precision cosmological data. We expect our formalism to produce similar results, although the simplicity of our equations may bring some computational benefits.

Winding back the evolution of the Universe to earlier times, the consequences of relativistic gravitational fields should be expected to become more significant. The corrections we found for the large-scale expansion of our particular solution, presented in Section \ref{sec6b}, were of the order of our cell size squared divided by the Hubble radius squared. Taking cells the size of the homogeneity scale, this gives corrections at the level of about $1$ part in $10^4$. Larger cells will give larger corrections. By coincidence, this is about the same size as the contribution of Cosmic Microwave Background (CMB) radiation to the expansion of the Universe at late-times. Our corrections also scale like a radiation fluid, so  naively extrapolating our results suggests that our corrections may start to become significant in the evolution of the Universe at about the same time as radiation starts to become important. For high precision observables, such as the CMB, it is conceivable that this could have some impact on the interpretation of data.

Let us now consider observations based on the evolution of null congruences in the space-time. The method of propagating null geodesics through a lattice universe of this type has already been considered in \cite{Tim1}. The method suggested there was to decompose the tangent vector to these curves as
\begin{equation}
k^a = (-u^b k_b) (u^a+n^a) \, ,
\end{equation}
where $u^a$ and $n^a$ are an orthogonal pair of time-like and space-like unit vectors. When a null geodesic reaches a cell boundary one can then read off $u^a k_a$ and $n^a$, which can be transformed to similar quantities in the new cell using the coordinate transformations from Eqs. \eqref{20a} and \eqref{20}. These quantities can then be used as initial conditions for extending the same null geodesic through the new cell.

Depending on the coordinate system being used, the cosmological redshift in these models comes primarily either from the expansion of space or the motion of the cell boundaries (for the time-dependent and time-independent backgrounds, respectively). As the leading order part of the expansion of each of our cells is given by a Friedmann-like equation with a dust-like content, we expect this to contribute the usual leading-order term to the redshifts calculated over large distances. What our framework adds beyond the usual treatment, however, is a consistent way to calculate the consequences of both Newtonian and post-Newtonian gravitational potentials on all scales up to the size of the cells being considered (larger scales can always be included using standard cosmological perturbation theory). The effect of relativistic gravity on galaxy redshifts has recently attracted much attention, in the light of upcoming surveys \cite{bonvin}.

Finally, the effect of inhomogeneity on the brightness and shear of a source can be calculated using the Sachs optical equations \cite{sachs}:
\begin{align}
\frac{d\theta}{d\lambda} +\theta^2 - \omega^2 + \sigma^* \sigma &= - \frac{1}{2} R_{ab} k^a k^b 
\nonumber \\
\frac{d \sigma}{d\lambda} +2 \sigma \theta &= C_{abcd} m^{a} k^b m^c k^d 
\nonumber \\
\frac{d\omega}{d\lambda} +2 \omega \theta &=0 \, ,
\end{align}
where $\theta$, $\sigma$ and $\omega$ are the expansion, shear, and vorticity scalars of the rays of light that are being modelled. The affine distance along these rays is denoted $\lambda$, and $R_{ab}$ and $C_{abcd}$ are the Ricci and Weyl tensors that describe the curvature of the space-time. The complex vectors $m^a$ are a set of mutually orthogonal and normalised vectors on the screen space orthogonal to $k^a$, and the angular diameter distance can be calculated by integrating the expansion scalar, such that $r_A \propto {\rm exp} (\int_e^o \theta d \lambda )$.

One can immediately notice that, when light propagates in the near-vacuum regions between galaxies, the driving term in the evolution equation for $\theta$ is absent, while the driving term in the corresponding equation for $\sigma$ is non-zero. This is exactly the opposite of what happens in homogeneous and isotropic space-times \cite{bert}. Our model provides a way in which this switching on and off of terms can be consistently modelled to post-Newtonian order in an expanding universe. Such quantities are of vital importance not only for constructing Hubble diagrams, but also for correctly interpreting data from galaxy surveys and the CMB. It is therefore important that the models used to interpret these observations are as robust as possible.
\vspace{2cm}

\section{Conclusions} \label{sec8}

We have constructed cosmological models by patching together many sub-horizon-sized regions of space, each of which is described using post-Newtonian physics. The boundaries between each of these regions was assumed to be reflection symmetric, in order to make the problem tractable. This allowed us to find the general form for the equation of motion of the boundary, as well as the general form of the post-Newtonian gravitational fields that arise for general matter content. These results both follow from a straightforward application of the junction conditions, which in the latter case provides the appropriate boundary conditions for solving Einstein's field equations.

As an example of how to apply this formalism, we considered the case of a large number of isolated masses, each of which is positioned at the centre of a cubic cell. We found that the large-scale evolution that emerges from such a configuration is well modelled by an equation that looks very much like the Friedmann equation of general relativity, with pressureless dust and radiation as sources. The radiation term is necessarily much smaller than the dust term (if the post-Newtonian expansion is to be valid), and appears in the Friedmann-like equation as if it had a negative energy density. This happens without any violation of the energy conditions, as the term in question arises from the non-linearity of Einstein's equations, and does not directly correspond to any matter content.

While small, the radiation-like term appears as the first relativistic correction to the large-scale expansion of the Universe, when the matter content is arranged in the way described above. This term provides a small negative contribution to the rate of expansion, and a small positive contribution to the rate of acceleration. The existence of a radiation-like term has been found previously using exact results derived for the evolution of reflection symmetric boundaries \cite{cgr},  and using the shortwave approximation for fluctuations around a background metric \cite{wald}.

It remains to be seen what form the first relativistic correction will take for more general configurations of matter fields, or for structures built with less restrictive symmetries. The former of these cases can be studied using the approach we have prescribed in Section \ref{sec6a}, while the latter requires a generalisation of our formalism.

\section*{Acknowledgements}

We are grateful to K. Malik and S. R\"{a}s\"{a}nen for helpful comments. VAAS and TC both acknowledge support from the STFC.

\appendix
\section{Geodesic Equation} \label{AppendixA}

The geodesic equation, in terms of proper time along a curve, is given by
\begin{align} 
\frac{d^2x^{a}}{d\tau^2} + \Gamma^{a}_{ \ bc} \frac{dx^{b}}{d\tau}  \frac{dx^{c}}{d\tau} = 0  \, , \label{A1}
\end{align}
where $\Gamma^{a}_{ \ bc}$ are the Christoffel symbols. Let us now consider the motion of a boundary at $x= X(t,y,z)$. The proper time derivatives of this boundary are given by
\begin{align}
\frac{dX}{d\tau} =& X_{,t} t_{,\tau} + X_{,A} x^{A}_{, \tau} \, , \label{A2} 
\end{align}
and
\begin{align}
\frac{d^2 X}{d\tau^2} =& X_{,tt} (t_{,\tau})^2 +  X_{,t} t_{,\tau\tau}+ 2X_{,tA} t_{,\tau} x^{A}_{, \tau} \nonumber \\
&+ X_{,AB} x^{A}_{,\tau} x^{B}_{, \tau} + X_{,A}x^{A}_{,\tau\tau}\, . \label{A3}
\end{align}
%where the last term in Eq. \eqref{A2} is $O(\epsilon^3)$, as $X_{,A} =O(\epsilon^2)$ at lowest order .  
Eq. \eqref{A2} then allows us to write the $t$, $y$ and $z$ components of the geodesic equation in terms of partial derivatives as
\begin{align} 
t_{,\tau\tau} &= - \Gamma^{t}_{ \ bc} (t_{,\tau})^2 x^{b}_{, t}  x^{c}_{, t} + O(\epsilon^5)  \ , \label{A4} \\
x_{,\tau\tau} &= - \Gamma^{x}_{ \ bc} (t_{,\tau})^2 x^{b}_{, t}  x^{c}_{, t} + O(\epsilon^6)  \ , \label{A5} \\
x^{A}_{,\tau\tau} &= - \Gamma^{A}_{ \ bc} (t_{,\tau})^2 x^{b}_{, t}  x^{c}_{, t} + O(\epsilon^4)  \, . \label{A6}
\end{align}
Using these equations, Eq. \eqref{A3} can be written as
\begin{align}
\frac{d^2 X}{d\tau^2} =&(t_{,\tau})^2 \bigg[X_{,tt}  - X_{,t} \Gamma^{t}_{ \ bc} x^{b}_{, t}  x^{c}_{, t}+ 2X_{,tA} x^{A}_{, t} \nonumber \\
&\hspace{1.2cm} + X_{,AB} x^{A}_{,t} x^{B}_{,t} - X_{,A} \Gamma^{A}_{ \ bc} x^{b}_{, t}  x^{c}_{, t}\bigg] \bigg|_{x=X}\, , \label{A7}
\end{align}
%The geodesic equation along the boundary is given by
%\begin{align} 
%&\frac{d^2 X}{d\tau^2} + \Gamma^{x}_{ \ bc} \frac{dx^{b}}{d\tau}  \frac{dx^{c}}{d\tau}\bigg|%_{x=X} = 0  \, . \label{A8}
%\end{align}
The Christoffel symbols required for evaluating Eq. \eqref{A1} can be simplified using Eq. \eqref{A7}, and written explicitly as
\begin{align}
%X_{,t}\Gamma^{t}_{\ bc} x^{b}_{, t}  x^{c}_{, t}  &= X_{,t}\Gamma^{t}_{ \ tt}+  {X_{,t}}^2\Gamma^{t}_{ \ xt} + O(\epsilon^6) \nonumber \\
\Gamma^{t}_{\ bc} x^{b}_{, t}  x^{c}_{, t} =& -\Phi_{,t} -\Phi_{,x}X_{,t} + O(\epsilon^5) \, , \label{A9} \\
%\Gamma^{A}_{\ bc} x^{b}_{, t}  x^{c}_{, t}  &= \Gamma^{A}_{ \ tt}+ O(\epsilon^4) \nonumber \\
\Gamma^{A}_{\ bc} x^{b}_{, t}  x^{c}_{, t} =& -\Phi_{,A} + O(\epsilon^4) \, , \label{A10} \\
%\Gamma^{x}_{\ bc} x^{b}_{, t}  x^{c}_{, t} &= \Gamma^{x}_{ \ tt} + 2\Gamma^{x}_{ \ tx}X_{,t}+ \Gamma^{x}_{ \ xx}{X_{,t}}^2 + O(\epsilon^6) \nonumber \\
\Gamma^{x}_{\ bc} x^{b}_{, t}  x^{c}_{, t} =& -\Phi_{,x} + 2\Phi\Phi_{,x} - \frac{h^{(4)}_{tt,x}}{2} + h^{(3)}_{ tx, t} \nonumber \\
&  + 2\Phi_{,t}X_{,t} + 2\Phi_{,x}{X_{,t}}^2 + O(\epsilon^6) \, , \label{A11}
\end{align}
where each term in these equations is taken to be evaluated on the boundary. 

Taking $x^{A}_{,t} = 0$, we can then see that Eqs. \eqref{A5}, \eqref{A7}, \eqref{A9}, \eqref{A10}, and \eqref{A11} allow the geodesic equation to be written as
%\begin{align} 
%& (t_{,\tau})^2 \bigg[X_{,tt} + X^{(2)}_{,A}\Phi_{,A} + \bigg(\Gamma^{x}_{bc} - \Gamma^{t}_{bc} X_{, t}\bigg)x^{b}_{, t} x^{c}_{, t}\bigg] \bigg|_{x=X} = 0 \ . \label{A13}
%\end{align}
%As $t_{,\tau} \neq 0$, and by evaluating the required Christoffel symbols, the equation of motion of the boundary is given by
\begin{align} \nonumber
X_{,tt} &= \bigg[ \Phi_{,x} - 2\Phi\Phi_{,x} + \frac{h^{(4)}_{tt,x}}{2} - h^{(3)}_{ tx, t} \\
 & \hspace{0.75cm} -3\Phi_{,x} X_{,t}^{2} -3 \Phi_{, t} X_{,t}  - X^{(2)}_{,A} \Phi_{,A} \bigg] \bigg|_{x=X} +  O(\epsilon^6) \, .
\end{align}
This is identical to Eq. \eqref{55}.

\section{Post-Newtonian Mass} \label{AppendixB}

In this subsection we will follow the approach used by Chandrasekhar \cite{Ch1, Ch2}.
If the 4-velocity is given by Eq. \eqref{33}, then the components of the energy-momentum tensor  are given by
\begin{align}
T^{ab} &= (\rho + \rho \Pi + p) u^{a} u^{b} + pg^{ab}    \label{B1} \, ,
\end{align}
such that
\begin{align}
T^{tt} &= \rho (1 +v^2+  \Pi + 2\Phi )  + O(\epsilon^6)   \nonumber \\ 
T^{t\mu} &= \rho \left(1 +v^2+  \Pi + 2\Phi+ \frac{p}{\rho} \right) v^{\mu}  + O(\epsilon^7)  \nonumber \, ,
\end{align}
and
\begin{align}  
T^{\mu \nu} &= \rho \left(1 +v^2+  \Pi + 2\Phi+ \frac{p}{\rho} \right) v^{\mu} v^{\nu} \nonumber \\&\quad + (1-2\Phi) p \delta^{\mu\nu}  + O(\epsilon^8) \ . \nonumber 
\end{align}
Let us now define $\sigma \equiv \rho(1+ v^2 + 2\Phi + \Pi + \frac{p}{\rho})$, and the total time derivative
\begin{align}
\frac{d}{dt}  \equiv \frac{\partial{}}{\partial{t}} + \bm{v} \cdot \nabla  \, .\label{B3}
\end{align}
To derive the form of the conserved post-Newtonian mass let us consider
\begin{align}
%T^{tb}_{\ \ ;b}  &= T^{tt}_{\ \ ,t} + \Gamma^{t}_{ct} T^{ct} + \Gamma^{t}_{ct} T^{tc} + T^{t\mu}_{\  \ ,\mu} + \Gamma^{t}_{c\mu} T^{c\mu} + \Gamma^{\mu}_{c\mu} T^{tc} =0\nonumber \\  \nonumber \\ 
%& = \frac{\partial}{\partial t} [\rho(1+ v^2 + 2\Phi + \Pi)]  +  \frac{\partial}{\partial x^{\mu}} [\rho v^{\mu} (1+ v^2 + 2\Phi + \Pi + \frac{p}{\rho})] + \rho \frac{\partial \Phi}{\partial t} \nonumber \\  \nonumber \\ 
T^{tb}_{\ \ ;b}  &= \sigma_{,t}  +  (\sigma v^{\mu})_{, \mu}  + \rho \Phi_{,t} -  p_{,t} \ . \label{B4}
\end{align}
Using the continuity equations, and Eq. \eqref{B3}, the last two terms in this equation can be written as
\begin{align}
\rho \Phi_{,t} -  p_{,t} &=  \rho \frac{d \Phi}{d t} - \frac{d p}{d t}  - v^{\mu} \bigg( \rho \Phi_{,\mu} -  p_{,\mu} \bigg)  \nonumber \\ 
%& = \rho \bigg( \frac{d \Phi}{d t}-v^{\mu} \frac{d v^{\mu}}{dt} \bigg) - \frac{d p}{d t} \nonumber \\ 
& =  \rho \bigg( \frac{d }{d t}(\Phi - \frac{1}{2} v^2) \bigg ) - \frac{d p}{d t}   \ .
\end{align}
Eq. \eqref{B3} can then be re-written as
\begin{align}
T^{tb}_{\ \ ;b} = \bigg(\frac{d }{d t} + \nabla \cdot \bm{v} \bigg)\sigma+ \rho \bigg( \frac{d }{d t} \left( \Phi - \frac{1}{2} v^2 \right) \bigg ) - \frac{d p}{d t}  \ . \label{B5}
\end{align}
We can now use the continuity equations to relate the time dependence of the post-Newtonian energy density to the pressure:
\begin{equation}
\rho \frac{d \Pi}{dt} = \frac{p}{\rho} \frac{d \rho}{d t} = - p \nabla \cdot \bm{v} \ . \label{B6}
\end{equation}
Thus, Eq. \eqref{B5} simplifies to
\begin{align}
%T^{tb}_{\ \ ;b} &= \bigg(\frac{d }{d t} + \nabla.v \bigg) \rho(1+ v^2 + 2\Phi + \Pi)+ p \nabla.v + \rho \bigg( \frac{d }{d t}(\Phi - \frac{1}{2} v^2) \bigg ) = 0 \nonumber \\ 
%&= \bigg(\frac{d }{d t} + \nabla.v \bigg) \rho(1+ v^2 + 2\Phi + \Pi)+ \rho \bigg( \frac{d }{d t}(\Phi - \frac{1}{2} v^2 - \Pi) \bigg ) = 0 \nonumber \\  \nonumber \\ 
%&= \bigg(\frac{d }{d t} + \nabla.v \bigg) \rho(1+ v^2 + 2\Phi + \Pi)+ \frac{d }{d t}\bigg[ \rho (\Phi - \frac{1}{2} v^2 - \Pi) \bigg] -  \frac{d\rho }{d t} \bigg((\Phi - \frac{1}{2} v^2 - \Pi) \bigg ) = 0\nonumber \\  \nonumber \\ 
%&= \bigg(\frac{d }{d t} + \nabla.v \bigg) \rho(1+ \frac{1}{2}v^2 + 3\Phi) +  \bigg(\frac{d\rho }{d t} +\rho \nabla.v \bigg)\bigg(\frac{1}{2} v^2 - \Phi + \Pi) \bigg ) = 0
%\nonumber \\  \nonumber \\ 
T^{tb}_{\ \ ;b}  &= \bigg(\frac{d }{d t} + \nabla \cdot \bm{v} \bigg) \rho \left( 1+ \frac{1}{2}v^2 + 3\Phi \right)  %   \nonumber \\ 
%&\equiv {\rho^{*}}_{,t}  + ( \rho^{*}v^{\mu})_{,\mu}   \, .
\end{align}
%where $\rho^{*}$ is a post-Newtonian energy density that satisfies the continuity equation. 
Hence, at $O(\epsilon^4)$, we can use the conservation of energy-momentum to identify the following conserved post-Newtonian mass, $M_{PN}$:
\begin{align}
M_{PN} &\equiv \int_{V} \rho \left( \frac{1}{2} v^2 + 3 \Phi \right)\ dV \nonumber \\ &= \frac{1}{2}\avg{\rho v^2} + 3\avg{ \rho\Phi} \, .
\end{align}


\begin{thebibliography}{99} 

\bibitem{mw} K. A. Malik and D. Wands, {\it Phys. Rep.} {\bf 475}, 1 (2009).

\bibitem{nbody} E. Bertschinger, {\it Ann. Rev. Astron. \& Astrophys.} {\bf 36 (1)}, 599 (1998); V. Springel {\it et al.}, {\it Nature} {\bf 435}, 629 (2005); A. Klypin, S. Trujillo-Gomez and J. Primack, {\it Astrophys. J.} {\bf 740}, 102 (2011).

\bibitem{br} C. Clarkson, G. Ellis, J. Larena and O. Umeh, {\it Rept. Prog. Phys.} {\bf 74}, 112901 (2011); T. Buchert and S. R\"{a}s\"{a}nen, {\it Ann. Rev. Nucl. Part. Sci.} {\bf 62}, 57 (2012); T. Clifton, {\it Int. J. Mod. Phys. D} {\bf 22}, 1330004 (2013).

\bibitem{clarkson} C. Clarkson and O. Umeh, {\it Class. Quant. Grav.} {\bf 28}, 164010 (2011).

\bibitem{rasanen} S. R\"{a}s\"{a}nen, {\it Phys. Rev. D} {\bf 81}, 103512 (2010).

\bibitem{euclid} http://sci.esa.int/euclid

\bibitem{ska} http://www.skatelescope.org/

\bibitem{bonvin} C. Bonvin, {\it Class. Quant. Grav.} {\bf 31}, 234002 (2014).

\bibitem{crt} T. Clifton, K. Rosquist and R. Tavakol, {\it Phys. Rev. D} {\bf 86}, 043506 (2012).

\bibitem{num} E. Bentivegna and M.  Korzynski, {\it Class. Quant. Grav.} {\bf 29}, 165007 (2012); E. Bentivegna, {\it Class. Quant. Grav.} {\bf 31}, 035004 (2014); E. Bentivegna and M.  Korzynski, {\it Class. Quant. Grav.} {\bf 30}, 235008 (2013).

\bibitem{num2} C.-M. Yoo, H. Abe, Y. Takamori and K.-i. Nakao, {\it Phys. Rev. D} {\bf 86}, 044027 (2012); C.-M. Yoo, H. Okawa and K.-i. Nakao, {\it Phys. Rev. Lett.} {\bf 111}, 161102 (2013); C.-M. Yoo and H. Okawa, {\it Phys. Rev. D} {\bf 89}, 123502 (2014).

\bibitem{pert} J.-P. Bruneton and J. Larena, {\it Class. Quant. Grav.} {\bf 29}, 155001 (2012).

\bibitem{Tim1} T. Clifton, {\it Class. Quant. Grav.} {\bf 28}, 164011 (2011).

\bibitem{LW} T. Clifton and P. G. Ferreira, {\it Phys. Rev. D} {\bf 80}, 103503 (2009); {\it Phys. Rev. D} {\bf 84}, 109902 (2011); T. Clifton and P. G. Ferreira, {\it JCAP} {\bf 0910}, 26 (2009).

\bibitem{cgrt} T. Clifton, D. Gregoris, K. Rosquist and R. Tavakol, {\it JCAP} {\bf 11}, 010 (2013).

\bibitem{cgr} T. Clifton, D. Gregoris and K. Rosquist, {\it Class. Quant. Grav.} {\bf 31}, 105012 (2014).

\bibitem{op1} T. Clifton, P. G. Ferreira and K. O'Donnell, {\it Phys. Rev. D} {\bf 85}, 023502 (2012).

\bibitem{op2} J.-P. Bruneton and J. Larena, {\it Class. Quant. Grav.} {\bf 30} 025002 (2013).

\bibitem{op3} R. G. Liu, arXiv:1501.05169 (2015).

\bibitem{kor1} M.  Korzynski, {\it Class. Quant. Grav.} {\bf 31}, 085002 (2014).

\bibitem{kor2} M.  Korzynski, arXiv:1412.3865 (2014).

\bibitem{cli1} T. Clifton, {\it Class. Quant. Grav.} {\bf 31}, 175010 (2014).

\bibitem{Will} C. M. Will, {\it Theory and experiment in gravitational physics}, Cambridge University Press (1993).

\bibitem{Is1} W. Israel, {\it Nuovo Cim. B}, \textbf{44}, 1 (1966).

\bibitem{volk} D. J. Croton {\it et al.}, {\it Mon. Not. R. Astron. Soc.} \textbf{365}, 11 (2005).

\bibitem{poly} H. S. M. Coxeter, {\it Regular polytopes}, Dover Publications (1973).

\bibitem{nbody2} J. Adamek, R. Durrer and M. Kunz, {\it Class. Quant. Grav.} {\bf 31}, 234006 (2014); J. Adamek, C. Clarkson, R. Durrer and M. Kunz, {\it Phys. Rev. Lett.} {\bf 114}, 051302 (2015); J. Adamek, D. Daverio, R. Durrer and M. Kunz, {\it Phys. Rev. D} {\bf 88}, 103527 (2013).

\bibitem{sachs} R. Sachs, {\it Proc. Roy. Soc. Lond. A} {\bf 264}, 309 (1961).

\bibitem{bert} B. Bertotti, {\it Proc. Roy. Soc. Lond. A} {\bf 294}, 195 (1966).

\bibitem{wald} S. R. Green and R. M. Wald, {\it Phys. Rev. D} {\bf 83}, 084020 (2011).

\bibitem{Ch1} S. Chandrasekhar, {\it Astrophys. J.} {\bf 158}, 45 (1969).

\bibitem{Ch2} S. Chandrasekhar, {\it Astrophys. J.} {\bf 142}, 1448 (1965).






\end{thebibliography}
\end{document}